\documentclass[twocolumn,superscriptaddress,pra]{revtex4-1}
\usepackage{graphicx}
\usepackage{amsmath}
\usepackage{amsthm}
\usepackage{amssymb}
\usepackage{latexsym}
\usepackage{array}
\usepackage{hyperref}
\usepackage{float}
\usepackage{amsfonts}
\usepackage{mathrsfs}
\usepackage{verbatim}
\usepackage{bbold}
\usepackage[normalem]{ulem}

\usepackage{tikz,pgfplots}
\pgfplotsset{compat=1.18}

\usepackage{upgreek}

\usepackage{makecell}
\usepackage{adjustbox,lipsum}

\usepackage{algorithm}
\usepackage{algpseudocode}

\usepackage{color}

\newcommand{\bra}[1]{\left<#1\right|}
\newcommand{\ket}[1]{\left|#1\right>}
\newcommand{\abs}[1]{\bigl|#1\bigr|}

\newcommand{\expt}[1]{\langle #1 \rangle}

\renewcommand\arraystretch{1.2}

\newtheorem{theorem}{Theorem}
\newtheorem{proposition}{Proposition}
\newtheorem{lemma}{Lemma}
\newtheorem{corollary}{Corollary}
\newtheorem{definition}{Definition}
\theoremstyle{remark}
\newtheorem{remark}{Remark}

\newcommand{\tr}{\mbox{Tr}}

\begin{document}

\title{Entangling Power and Its Deviation: A Quantitative Analysis on Input-State Dependence and Variability in Entanglement Generation}

\author{Kyoungho~Cho}
\affiliation{Department of Statistics and Data Science, Yonsei University, Seoul 03722, Republic of Korea}
\affiliation{Institute for Convergence Research and Education in Advanced Technology, Yonsei University, Seoul 03722, Republic of Korea}

\author{Jeongho~Bang}\email{jbang@yonsei.ac.kr}
\affiliation{Institute for Convergence Research and Education in Advanced Technology, Yonsei University, Seoul 03722, Republic of Korea}

\date{\today}

\date{\today}% It is always \today, today, but any date may be explicitly specified

\begin{abstract}
Quantifying the entangling capability of quantum operations is a fundamental task in quantum information science. Traditionally, this capability is measured by the entangling power (EP), defined as the average entanglement generated when a quantum operation acts uniformly on all possible product states. However, EP alone cannot capture the intricate input-state-dependent nature of entanglement generation. To address this, we define a complementary metric---entangling power deviation (EPD)---as the standard deviation of entanglement generated over all product input states, thereby capturing the multifaceted nature of entangling behavior. We develop a general group-theoretical framework that yields closed-form expressions for both EP and EPD. Our analysis shows that any nontrivial entangling operation necessarily exhibits input-state dependence: nonzero EP implies a nonzero EPD. By analyzing representative two-qubit gates, we show that the gates with identical EP can exhibit markedly different EPD values, illustrating that the nature of entanglement generation can significantly differ depending on the gate functionality. Extending our framework to a class of generalized controlled-unitary operations acting on bipartite Hilbert spaces of arbitrary dimensions, we further analyze the interplay between the entangling strength and uniformity, as quantified by EP and EPD. Moreover, we uncover a subtle dimension-parity-dependent behavior in entanglement generation, which EP alone fails to detect. These findings highlight EPD as an indispensable diagnostic tool---one that, alongside EP, provides a deeper and more complete characterization of the entangling structure.
\end{abstract}

%\keywords{Quantum computation, Variational Quantum Eigensolver(VQE), Quantum amplitude amplification}
\maketitle

%-------------------------------------------------------------------------------------------------------------------------------------------------------------------------------------------------------------------------------------
\section{Introduction}\label{Sec:1}
%-------------------------------------------------------------------------------------------------------------------------------------------------------------------------------------------------------------------------------------

Entanglement lies at the heart of quantum theory, representing one of its most profound and genuinely nonclassical phenomena. Over the past few decades, it has been recognized not merely as a theoretical interest, but as a physical resource---a foundational enabler of quantum information processing. As such, it plays a central role in a wide range of quantum technologies, for example, quantum teleportation and quantum secure communication. In particular, the ability to entangle qubits is essential to quantum error correction, variational quantum algorithms, and other protocols that exploit quantum correlations to encode and process information beyond classical reach. Accordingly, it is crucial to assess and quantify how effectively a given quantum operation can generate entanglement. This motivates the development of rigorous and generalizable metrics that characterize the entangling capability of quantum gates and channels. Such quantification not only deepens our understanding of the structural role of entanglement but also informs the design or benchmarking of quantum circuits.

A standard approach to quantify the quantum entanglement is the entangling power (EP), which measures the average amount of entanglement generated when a given quantum operation acts on uniformly distributed product input states. First introduced by Zanardi et al.~\cite{PhysRevA.62.030301}, EP serves as a statistical characterization of the entangling capability of unitary operators. Since its introduction, EP has been extensively studied (see Refs.~\cite{PhysRevA.63.040304,PhysRevA.95.040302, 2410.03361, PhysRevA.110.062422,PhysRevResearch.2.043126,Pal:2018nvj, J.Batle:2005,Padmanabhan:2020hfj}) and linked to a variety of key topics, including operator entanglement~\cite{PhysRevA.66.044303,PhysRevA.110.052416,PhysRevA.83.062320,PhysRevA.67.042323,1204.0996}, quantum circuit complexity~\cite{PhysRevLett.127.020501, Azado:2024iam}, and indicators of quantum chaos~\cite{PhysRevE.70.016217,arXiv:0504052}. Its conceptual simplicity and analytic tractability make EP a widely used and intuitive tool for assessing the first-order entangling behavior of quantum operations. However, by construction, EP captures only the average nature of entanglement generation over all input states, and thus cannot distinguish between the quantum operations that produce entanglement uniformly and those that exhibit strong dependence on the inputs; hence, EP alone offers an incomplete picture of the entanglement generation. This limitation would become especially critical in practical settings where the performances are highly sensitive to input-state variations.

To overcome this limitation, we introduce a complementary metric: the entangling power deviation (EPD). Defined as the standard deviation of entanglement generated over all product input states, EPD quantifies the extent to which the entangling capability of a quantum operation depends on the input states. This EPD, originally proposed in the present study, provides a second-order characterization that captures the inherent input-state dependence in entanglement generation---revealing structural features that remain hidden under average-based measure EP. This added layer of characterization proves essential in both theoretical analyses and practical applications, such as quantum circuit design, quantum error mitigation, and benchmarking of noisy quantum devices.

In this paper, we establish a general group-theoretical framework for computing EP and EPD of unitary operations acting on arbitrary-dimensional bipartite Hilbert spaces. Whereas Ref.~\cite{PhysRevA.62.030301} evaluates EP using the second Haar moment of random product inputs (i.e., the two-copy moment operator), EPD defined here necessarily involves the fourth moment. Accordingly, we develop a systematic group-theoretical moment-operator framework that characterizes the general $\kappa$th moments and reduces the resulting $\kappa$-copy trace expressions, making the $\kappa=4$ analysis tractable and providing a general tool beyond the EP-only treatment in Ref.~\cite{PhysRevA.62.030301}. This formalism thus yields closed-form expressions for both quantities and offers physical insight into the intrinsic structure underlying the entanglement generation. Notably, we show that perfectly uniform entanglement generation is impossible within this setting: any nonzero EP necessarily accompanies a nonzero EPD. This provides a rigorous link between entangling strength and input-state dependence that is invisible to EP alone. We then analyze several well-known two-qubit gates and find that the gates with identical EP can exhibit significantly different EPD values. This observation highlight the necessity of EPD for differentiating the quantum operations that appear equivalent under average-based assessments but diverge in their entanglement fluctuation profiles. We further extend our analysis to a class of generalized controlled-unitary (CU) operations acting on arbitrary-dimensional bipartite systems. Herein, we further analyze how the entangling strength and its input-state variability co-vary in arbitrary dimensions, as quantified by EP and EPD. Furthermore, we unveil a delicate dimension-parity-dependent structure in the high-dimensional entangling landscape---one that eludes the detection by EP alone and only becomes apparent through the lens of EPD. The proposed framework and findings not only deepen our theoretical understanding but also provide practical diagnostic tools to characterize the entangling nature. In particular, EPD offers a more complete perspectives for the design of entangling operations, e.g., in quantum error correction~\cite{Bennett1996mixed,Scott2004multipartite} or quantum machine learning~\cite{Wiersema2020exploring,Eisert2021entangling,Azad2023qleet}.

%-------------------------------------------------------------------------------------------------------------------------------------------------------------------------------------------------------------------------------------
\section{Entangling Power and Its Deviation}\label{Sec:2}
%-------------------------------------------------------------------------------------------------------------------------------------------------------------------------------------------------------------------------------------

To utilize quantum operations effectively, it is essential to quantitatively understand and evaluate how well a given quantum operation can generate entanglement. A natural and practical starting point for this task is to measure how effectively a quantum operation can transform a product state into an entangled one. In this context, we consider the entanglement properties of a general bipartite quantum system composed of two subsystems, denoted as ``subsystem $1$'' and ``subsystem $2$.'' Herein, each subsystem is associated with a Hilbert-space denoted as $\mathcal{H}_1$ and $\mathcal{H}_2$ respectively, and the total system is represented as $\mathcal{H}_{12}$. The quantum operation under consideration is a unitary operator $\hat{U}$ acting on the full space $\mathcal{H}_{12}$. Given an arbitrary pure product state $\ket{\psi_1} \otimes \ket{\psi_2}$, the output state after applying $\hat{U}$ is a new pure state,
\begin{eqnarray}
\ket{\Psi_{12}} = \hat{U}\ket{\psi_1} \otimes \ket{\psi_2}.
\label{eq:Psi_12}
\end{eqnarray}
The degree of entanglement in this output state can be quantified in various ways. Throughout this work, we adopt the linearized entropy as the entanglement measure due to its computational simplicity and physical interpretability. The linearized entropy is defined as~\cite{Buscemi2007linear}
\begin{eqnarray}
E(\ket{\Psi_{12}}) = 1 - \tr \hat{\rho}_j^2~(j=1~\text{or}~2),
\label{eq:linear_E}
\end{eqnarray}
where $\hat{\rho}_j$ is a reduced density matrix of the subsystem $j$, obtained by tracing out another subsystem. This quantity takes the value zero for separable states and reaches its maximum for maximally entangled states.

The entangling power of a unitary operation $\hat{U}$, denoted by $e_p(\hat{U})$, is then defined as the average amount of the entanglement generated when $\hat{U}$ is applied to all possible product states~\cite{PhysRevA.62.030301, PhysRevA.63.040304}:
\begin{eqnarray}
e_p(\hat{U}) = \langle E(\ket{\Psi_{12}}) \rangle = \int d \mu(\psi_1, \psi_2) E(\hat{U} \ket{\psi_1} \otimes \ket{\psi_2}),
\label{eq:EP} 
\end{eqnarray}
where $\mu(\psi_1,\psi_2)$ denotes the Haar measure over product states  $\ket{\psi_1}\otimes\ket{\psi_2}$. EP is the average of the linearized entropy $E$ over the entire space of input product states. It serves as a statistical indicator especially useful when the input states are unknown or variable. EP has demonstrated significant practical utility in various areas of quantum information science. For instance, in the design and benchmarking of quantum circuit, EP serves as a tool for comparing the entanglement-generating capability of different entangling quantum gates. Since EP does not depend on the computational basis or specific input states, it provides a fair measure of a gate's general entangling ability. In fact, EP has been employed in the classification of two-qubit gates to distinguish between local and nonlocal gates, and to identify those that generate maximal average entanglement~(see Refs.~\cite{PhysRevResearch.2.043126,PhysRevA.70.052313} for more details). EP also finds applications in analyzing quantum circuit complexity, where it serves as a proxy for the expressive power of parameterized quantum circuits (PQCs)~\cite{Eisert2021entangling}. In variational quantum algorithms, PQCs with higher EP are often more effective in exploring complex entangled states~\cite{Sim2019expressibility}. Furthermore, EP plays a role in quantum error correction; operations with high EP in encoding and recovery procedures may indicate the presence of strong quantum correlations necessary for protecting logical information against noise~\cite{Bennett1996mixed}.

However, as an average quantity, EP provides no information about how the amount of entanglement varies across different input states. Specifically, two unitary operations may have identical EP values, yet one may generate entanglement uniformly across all inputs, while the other only entangles a few specific states strongly and most others weakly or not at all. To capture this distinction and address the subtle limitations of EP, we introduce an additional quantity, named as entangling power deviation (EPD). EPD quantifies how broadly the entanglement values are distributed over the space of input product states. In other words, it is defined as the standard deviation of the entanglement values:
\begin{eqnarray} 
\Delta_p(\hat{U}) = \sqrt{ \langle E^2(\ket{\Psi_{12}}) \rangle - e_p(\hat{U})^2}. \label{defDp}
\end{eqnarray}
A small value of EPD implies that the unitary operation generates entanglement at a consistent level regardless of input, while a large EPD indicates that the amount of entanglement depends strongly on the input state. EPD provides information that cannot be captured by EP alone. By quantifying the variance in entanglement generation over all input states, EPD reveals how uniformly a unitary operation acts as an entangler. It thus provides critical insight into the consistency and input-state sensitivity of the entanglement production. In particular, even among operations with identical EP values, differences in EPD can highlight structural differences: some operations entangle all input states uniformly, while others concentrate entangling capacity on a narrow subset of inputs.

Ultimately, EP and EPD together offer a richer and more precise framework for evaluating the entangling behavior of quantum operations. While EP characterizes average capability, EPD reveals the reliability and consistency of the capability. These complementary indicators can be used as practical tools for selecting and optimizing quantum gates in a wide range of applications, including quantum algorithm design, quantum error correction, and quantum neural network architectures. This approach, which goes beyond the limitations of a single average quantity, enables a deeper understanding of entangling behavior in quantum systems.

However, it should also be noted that while the pair of EP and  EPD provides a concise summary of typical entangling strength and its input-state sensitivity, it does not uniquely determine the full shape of the probability distribution, say $P_{\hat{U}}(e)$, of the entanglement values $e:=E(\ket{\Psi_{12}})$ (e.g., skewness, multimodality, or how much probability mass accumulates near the edges of its support); in particular, since $e$ is bounded, $P_{\hat{U}}(e)$ cannot be heavy-tailed in the strict sense even though it can still be strongly skewed or sharply peaked.

%-------------------------------------------------------------------------------------------------------------------------------------------------------------------------------------------------------------------------------------
\section{Group-Theoretic Analysis of Entangling Power and Entangling Power Deviation}\label{Sec:3}
%-------------------------------------------------------------------------------------------------------------------------------------------------------------------------------------------------------------------------------------

We investigate EP and EPD of a quantum operation acting on bipartite quantum systems building upon the group-theoretic formalism. Specifically, we derive the analytical expressions for EPD and EPD by exploiting the properties of Haar measure and Schur-Weyl duality~\cite{goodman2000representations}. This approach allows us to decompose multiple-copy Hilbert spaces into symmetric and antisymmetric subspaces, greatly simplifying the Haar integrals. Consequently, both the average and the variance of entanglement generation can be analyzed compactly in terms of permutation operators and symmetric projectors.

\subsection{Preliminary: Permutation operators and projectors} %-------------------------------------------------------------------------------

%We first present general formulae valid for arbitrary distributions over product states, and then specialize to the case of a uniform Haar distribution. Throughout this section, we use the shorthand notation $\tr_{1234}$ and $\tr_{12345678}$ to denote partial traces over the tensor product spaces $(\mathcal{H}_1 \otimes \mathcal{H}_2)^{\otimes 2}$ and $(\mathcal{H}_1 \otimes \mathcal{H}_2)^{\otimes 4}$, respectively. In the 2-copy space, the subsystems $\mathcal{H}_1$ and $\mathcal{H}_2$ correspond to indices $1,3$ and $2,4$, respectively; in the 4-copy space, to $1,3,5,7$ and $2,4,6,8$, respectively. We denote by $\mathcal{B}(\mathcal{H})$ the algebra of bounded linear operators on a Hilbert space $\mathcal{H}$, and by $\mathcal{L}(\mathcal{H})$ the set of linear operators on $\mathcal{H}$.

Firstly, we introduce the permutation operators associated with symmetric groups, which play a central role in the presented framework. The permutation operators are crucial for characterizing the symmetries of tensor product spaces, since they arise naturally in the evaluation of Haar integrals over multiple copies of quantum systems. In particular, these operators enable us to express the projectors onto symmetric or anti-symmetric subspaces.

The relevant mathematical structure which is dealt with here is the symmetric group. Thus, we denote as follows: the symmetric group of degree $\kappa$ (i.e., all permutations of $\kappa$ elements) is denoted by $S_\kappa$, and for a finite set $X$, its symmetric group is denoted by $S_X$. These notations will be used throughout this work. Their relevance to Haar integration and representation theory is well described in Ref.~\cite{Zhang:2014zbq,Ragone:2022axl,Mele2024introductiontohaar} (or see {\bf Appendix~\ref{appendix:A}}). Now, we define the permutation operator acting on $\kappa$-fold tensor product spaces:
\begin{definition}\label{def:1}
Let $S_\kappa$ be the symmetric group of degree $\kappa$. For $\pi \in S_\kappa$, the corresponding permutation operator $\hat{V}_d(\pi)$ on $(\mathbb{C}^d)^{\otimes \kappa}$ is defined by
\begin{eqnarray}
\hat{V}_d(\pi) = \sum_{j_1, \cdots, j_\kappa = 0}^{d-1} \ket{j_{\pi^{-1}(1)}, \cdots, j_{\pi^{-1}(\kappa)}}\bra{j_1, \cdots, j_\kappa}.
\label{eq:defVd}
\end{eqnarray}
\end{definition}
This operator acts by permuting the tensor factors of basis elements, and serves as a building block for defining symmetries and projecting onto irreducible components in multiple-copy Hilbert spaces. In particular, for the $2$-copy and $4$-copy spaces of bipartite systems, we represent the permutation operators $\hat{\xi}_\pi$ for $\pi \in S_4$ and $\hat{\chi}_\nu$ for $\nu \in S_8$, such that
\begin{eqnarray}
\hat{\xi}_\pi &=& \sum_{j_1, \cdots, j_4=0}^{d-1} \ket{j_{\pi^{-1}(1)}, \cdots, j_{\pi^{-1}(4)}}\bra{j_1, \cdots, j_4 }, \\
\hat{\chi}_\nu &=& \sum_{j_1, \cdots, j_8=0}^{d-1} \ket{j_{\nu^{-1}(1)}, \cdots, j_{\nu^{-1}(8)}} \bra{j_1, \cdots, j_8},
\end{eqnarray}
where $\hat{\xi}_\pi$ and $\hat{\chi}_\nu$ can be written by $d^4 \times d^4$ and $d^{8} \times d^{8}$ matrices over $\mathbb{C}$, respectively. 
%In the context of entangling power and its deviation, such permutation operators allow for compact representations of moments of entanglement over product state ensembles. 

Here, we provide the useful properties of the permutation operators as follows. Let $\pi, \nu \in S_\kappa$. The permutation operator $\hat{V}_d(\pi)$ acting on $(\mathbb{C}^d)^{\otimes \kappa}$ satisfies:
\begin{widetext}
\begin{eqnarray}
%\begin{array}{ll}
&&(i)~\hat{V}_d(id) = \hat{\openone}, \nonumber \\
&&(ii)~\hat{V}_d(\pi) \hat{V}_d(\nu) = \hat{V}_d(\pi \circ \nu), \nonumber \\
&&(iii)~\hat{V}_d(\pi)^\dagger = \hat{V}_d(\pi^{-1}), \nonumber \\
&&(iv)~\hat{V}_d(\pi) \ket{\psi_1} \otimes \cdots \otimes \ket{\psi_\kappa} = \ket{\psi_{\pi^{-1}(1)}} \otimes \cdots \otimes \ket{\psi_{\pi^{-1}(\kappa)}}~\text{for product states $|\psi_j\rangle \in \mathbb{C}^d$}, \nonumber \\
&&(v)~\hat{V}_d(\pi) \bigl(\hat{A}_1 \otimes \cdots \otimes \hat{A}_\kappa \bigr) \hat{V}_d(\pi)^\dagger = \hat{A}_{\pi^{-1}(1)} \otimes \cdots \otimes \hat{A}_{\pi^{-1}(\kappa)}~\text{for operators $\hat{A}_j \in \mathcal{L}(\mathbb{C}^d)$},
\label{eq:propVd}
%\end{array}
\end{eqnarray}
\end{widetext}
where $id \in S_\kappa$ is an identity cycles, $\hat{\openone}$ is an identity matrix, and $\mathcal{L}(\mathcal{H})$ denotes the set of linear operators on $\mathcal{H}$. These properties follow directly from the definition of $\hat{V}_d(\pi)$ as a permutation operator acting on tensor product spaces. Each statement can be verified by straightforward substitution and computation.

The properties (i), (ii), and (iii) imply that the set of operators $\{ \hat{V}_d(\pi) \}_{\pi \in S_\kappa}$ forms a unitary representation of the symmetric group $S_{\kappa}$ on $(\mathbb{C}^d)^{\otimes \kappa}$. This representation acts by permuting the tensor factors of the product states and operators according to the inverse permutation, as described in (iv) and (v). In particular, property (v) implies that for an operator $\hat{A}$ acting on $\kappa$ Hilbert spaces, denoted by $\hat{A}_{1 \cdots \kappa}$, the conjugation action of $\hat{V}_d(\pi)$ results in the reordering of the subsystems:
\begin{eqnarray}
\hat{V}_d(\pi) \hat{A}_{1 \cdots \kappa} \hat{V}_d(\pi)^{-1} = \hat{A}_{\pi^{-1}(1) \cdots \pi^{-1}(\kappa)}.
\label{eq:conjVdA}
\end{eqnarray}
For simplicity, from now on we rewrite this as 
\begin{eqnarray}
\hat{V}_d(\pi) \hat{A} \hat{V}_d(\pi)^{\dagger} = \hat{A}_{\pi^{-1}}.
\label{eq:conjVdA2}
\end{eqnarray}
These permutation operators serve as essential tools for constructing (anti-)symmetric subspaces, specifically by enabling projections onto the irreducible components of $(\mathbb{C}^d)^{\otimes \kappa}$ via Schur–Weyl duality. In this context, the representation $\{\hat{V}_d(\pi)\}_{\pi \in S_{\kappa}}$ generates the commutant of the collective $U(d)$ action on the $\kappa$-fold tensor product space.

We then consider the subspace of states invariant under arbitrary permutations, which is known as the symmetric subspace
%\begin{widetext}
\begin{definition}[Symmetric subspace]
Symmetric subspace is defined as: For all $\pi \in S_\kappa$,
\begin{eqnarray}
\text{Sym}_{\kappa}(\mathbb{C}^{d})= \left\{ \ket{\psi} \in (\mathbb{C}^d)^{\otimes \kappa}: \hat{V}_d(\pi) \ket{\psi} = \ket{\psi} \right\}.
\end{eqnarray}
\end{definition}
%\end{widetext}
This symmetric subspace is governed by the operator
\begin{eqnarray}
\hat{P}^{(d,\kappa)}_{\text{sym}} = \frac{1}{\kappa !} \sum_{\pi \in S_\kappa} \hat{V}_d(\pi),
\label{eq:proj_sym}
\end{eqnarray}
satisfying $\hat{V}_d(\pi) \hat{P}_{\textrm{sym}}^{(d,\kappa)} = \hat{P}_{\textrm{sym}}^{(d,\kappa)}$, $\hat{P}_{\textrm{sym}}^{(d,\kappa)2} = \hat{P}_{\textrm{sym}}^{(d,\kappa)}$, and $\hat{P}_{\textrm{sym}}^{(d,\kappa) \dagger} = \hat{P}_{\textrm{sym}}^{(d,\kappa)}$. Therefore, $\hat{P}_{\textrm{sym}}^{(d,\kappa)}$ is an orthogonal projector, and indeed $\textrm{Sym}_{\kappa}(\mathbb{C}^{d}) = \operatorname{Im}(P_{sym}^{(d,\kappa)})$.

Conversely, we can also define anti-symmetric subspace, consisting of the states that pick up a sign factor under permutations.
%\begin{widetext}
\begin{definition}[Anti-symmetric subspace]
Anti-symmetric subspace is defined as: For all $\pi \in S_\kappa$,
\begin{eqnarray}
\text{ASym}_{\kappa}(\mathbb{C}^{d}) \!=\! \left\{\! \ket{\psi} \!\in\! (\mathbb{C}^d)^{\otimes \kappa} \!:\! \hat{V}_d(\pi)\!\ket{\psi} \!=\! \text{sgn}(\pi)\!\ket{\psi} \!\right\},
\end{eqnarray}
where $\textrm{sgn}(\pi)$ denotes the sign of a permutation $\pi$.
\end{definition}
%\end{widetext}
The anti-symmetric subspace is governed by 
\begin{eqnarray}
\hat{P}^{(d,\kappa)}_{\textrm{asym}} = \frac{1}{\kappa !} \sum_{\pi \in S_\kappa} \text{sgn}(\pi) \hat{V}_d(\pi),
\label{eq:proj_asym}
\end{eqnarray}
satisfying $\hat{V}_d(\pi) \hat{P}_{\textrm{asym}}^{(d,\kappa)} = \text{sgn}(\pi) \hat{P}_{\textrm{asym}}^{(d,\kappa)}$, $\hat{P}_{\textrm{asym}}^{(d,\kappa)2} = \hat{P}_{\textrm{asym}}^{(d,\kappa)}$, and $\hat{P}_{\textrm{asym}}^{(d,\kappa) \dagger} = \hat{P}_{\textrm{asym}}^{(d,\kappa)}$. Therefore, $\hat{P}_{\textrm{asym}}^{(d,\kappa)}$ is an orthogonal projector, and $\text{ASym}_{\kappa}(\mathbb{C}^{d}) = \operatorname{Im}(P_{asym}^{(d,\kappa)})$. We note that the projectors ${P}_{\textrm{asym}}^{(d,\kappa)}$ and $\hat{P}_{\textrm{sym}}^{(d,\kappa)}$, given by Eq.~(\ref{eq:proj_sym}) and Eq.~(\ref{eq:proj_asym}), exhibit the orthogonality, i.e., 
\begin{eqnarray}
\hat{P}_{\textrm{asym}}^{(d,\kappa) \dagger} \hat{P}_{\textrm{sym}}^{(d,\kappa)}=0,
\label{eq:orth_P}
\end{eqnarray}
which are useful when decomposing the full Hilbert space into invariant subspaces.

In the context of quantities such as EP and EPD, it is frequently necessary to symmetrize only over a subset of subsystems. This leads us to the notion of partially symmetrized and anti-symmetrized projectors.
\begin{definition}[Partially symmetrized and anti-symmetrized projectors]
Given a subset $X \subseteq \{1,2,\cdots, \kappa \}$ of subsystems, the partially symmetrized and anti-symmetrized projectors are defined, respectively, as
\begin{eqnarray}
\hat{P}_X^{+} &=& \frac{1}{|X|!} \sum_{\pi \in S_X} \hat{V}_d(\pi), \\
\hat{P}_X^{-} &=& \frac{1}{|X|!} \sum_{\pi \in S_X} \text{sgn}(\pi) \hat{V}_d(\pi),
\end{eqnarray}
where $S_X$ is the symmetric group acting on the positions in $X$, leaving the remaining subsystems fixed and $\hat{V}_d$ is the permutation matrices over the whole system.
\label{def:proj}
\end{definition}
Because $S_X$ is the subset of the whole system $S_{\kappa}$, $\pi \in S_X$ are also cycles in $\pi \in S_{\kappa}$; i.e., $S_X$ is a subgroup of $S_{\kappa}$. Thus, for the permutations matrices over partial symmetrized and anti-symmetrized matrices are written by the permutation matrices on the total system.

In the cases of $2$-copy and $4$-copy spaces, the projectors take the following explicit forms: For any specific $j, k$ in $2$-copy spaces and $j,k,l,m$ in $4$ copy spaces, $\hat{P}_{jk}^{\pm} = \frac{1}{2} \bigl( \hat{\openone} \pm \hat{T}_{jk} \bigr)$, $\hat{P}_{jklm}^{+} = \frac{1}{4!} \sum_{\pi \in S_4} \hat{\xi}_\pi$, and $\hat{P}_{jklm}^{-} = \frac{1}{4!} \sum_{\pi \in S_4} \textrm{sgn}(\pi) \hat{\xi}_\pi$, where $\hat{T}_{jk}$ is the swap between the subsystems 1st and 3rd while leaving others unchanged. Here, $S_4 = \{ j, k, l, m\}$. These partial symmetrized and anti-symmetrized projectors play a pivotal role in our group-theoretic formalism. Specifically, they enable efficient decomposition of multiple-copy Hilbert spaces and simplify the evaluation of the Haar integrals in the computation of EP and EPD.

\subsection{Analysis of EP and EPD based on subspace permutation symmetry} %-------------------------------------------------------------------------------

Within a group-theoretical framework, we provide an explicit representation of EP and EPD. Most previous studies have assumed uniformly distributed input product states to express the entangling capability of a target unitary $\hat{U}$ solely in terms of its intrinsic elements~\cite{PhysRevA.62.030301}. Thus, we present the following proposition:
\begin{widetext}
\begin{proposition}\label{prop:1}
For the uniform distribution $\overline{p}_0$ over the input product states $\ket{\Phi_{12,\mathrm{in}}} = \ket{\psi_1} \otimes \ket{\psi_2} \in {\mathcal H}_1 \otimes {\mathcal H}_2$, EP and EPD are given by:
\begin{eqnarray}
e_{\overline{p}_0}(\hat{U}) &=& 2 \tr \bigl( \hat{U}^{\otimes 2} \hat{\Omega}_{\overline{p}_0}^{(2)} \hat{U}^{\dagger \otimes 2} \hat{P}_{13}^{-} \bigr), \label{eq:EP_explicit_p0} \\
\Delta_{\overline{p}_0}(\hat{U}) &=& \sqrt{4 \tr \bigl( \hat{U}^{\otimes 4} \hat{\Omega}_{\overline{p}_0}^{(4)} \hat{U}^{\dagger \otimes  4} \hat{P}_{13}^{-}  \hat{P}_{57}^{-} \bigr) - e_{\overline{p}_0}(\hat{U})^2}, \label{eq:EPD_explicit_p0}
\end{eqnarray}
where the operator $\hat{\Omega}_{\overline{p}_0}^{(\kappa)}$ represents the ensemble-averaged product states, defined as
\begin{eqnarray}
\hat{\Omega}_{\overline{p}_0}^{(\kappa)} = \int d \mu(\psi_1, \psi_2) \ket{\Phi_{12,\mathrm{in}}}\bra{\Phi_{12,\mathrm{in}}}^{\otimes \kappa}.
%\hat{\Omega}_p^{(\kappa)} = \int d \mu(\psi_1, \psi_2) \left( \ket{\psi_1}\bra{\psi_1} \otimes \ket{\psi_2}\bra{\psi_2}\right)^{\otimes \kappa}, 
%\hat{\Omega}_p^{(4)} &=& \int d \mu(\psi_1, \psi_2) \left( \ket{\psi_1}\bra{\psi_1} \otimes \ket{\psi_2}\bra{\psi_2}\right)^{\otimes 4},
\label{eq:Omega_k}
\end{eqnarray}
Here, we assign the indices of the subsystems from $1$ to $2\kappa$ by rewriting the states of the copies, such that
\begin{eqnarray}
\ket{\Phi_{12,\mathrm{in}}}\bra{\Phi_{12,\mathrm{in}}}^{\otimes \kappa} &=& \bigl(\ket{\psi_1}\bra{\psi_1} \otimes \ket{\psi_2}\bra{\psi_2} \bigr)^{\otimes \kappa} \nonumber \\
  &=& \ket{\psi_1}\bra{\psi_1} \otimes \ket{\psi_2}\bra{\psi_2} \otimes \cdots \otimes \ket{\psi_{2\kappa}}\bra{\psi_{2\kappa}},
\label{eq:re_indices}
\end{eqnarray}
where the states $\ket{\psi_{j \in {\mathcal I}_\text{odd}}}$ and $\ket{\psi_{j \in {\mathcal I}_\text{even}}}$ are the copies of $\ket{\psi_1} \in {\mathcal H}_1$ and $\ket{\psi_2} \in {\mathcal H}_2$, respectively. Note that ${\mathcal I}_\text{odd} =\{1,3,\ldots,2\kappa-1\}$ and ${\mathcal I}_\text{even} =\{2,4,\ldots,2\kappa\}$. Then, the states $\hat{\Omega}_{\overline{p}_0}^{(2)}$ and $\hat{\Omega}_{\overline{p}_0}^{(4)}$ are given by
\begin{eqnarray}
\hat{\Omega}_{\overline{p}_0}^{(2)} = (2!)^2 C_{d_1} C_{d_2} \hat{P}_{13}^{+} \hat{P}_{24}^{+}, \quad \hat{\Omega}_{\overline{p}_0}^{(4)} = (4!)^2 D_{d_1} D_{d_2} \hat{P}_{1357}^{+} \hat{P}_{2468}^{+}.
\label{eq:Ometa_prop2}
\end{eqnarray}
where the subscripts $1, 3, 5, 7$ and $2, 4, 6, 8$ denote the subsystem indices associated with the copies of the states in $\mathcal{H}_1$ and $\mathcal{H}_2$, respectively. The projectors $\hat{P}_{ij}^{+}$ and $\hat{P}_{ijkl}^{+}$ performs the projections onto the symmetric subspaces of the corresponding subsystems. The constants $C_{d_j}$ and $D_{d_j}$ are
\begin{eqnarray}
C_{d_j} = \frac{1}{d_j (d_j + 1)}, \quad D_{d_j} = \frac{1}{d_j (d_j + 1) (d_j + 2) (d_j + 3)},
\label{eq:CDdj}
\end{eqnarray}
where $d_j = \dim(\mathcal{H}_j)~(j=1,2)$.
\end{proposition}
\end{widetext}

%denotes the measure on $\psi_1, \psi_2$ under $p(\psi_1, \psi_2)$.
\begin{proof}---We begin by recalling the ensemble-averaged product state $\hat{\Omega}_{\overline{p}_0}^{(\kappa)}$ in Eq.~(\ref{eq:Omega_k}). By noting that we have a product measure $\mu(\psi_1,\psi_2) = \mu(\psi_1) \mu(\psi_2)$ for the uniform distribution $\overline{p}_0$, we can rewrite Eq.~(\ref{eq:Omega_k}) as
\begin{eqnarray}
%\hat{\Omega}_{\overline{p}_0}^{(\kappa)} = \int d\mu(\psi_1, \psi_2) \ket{\Phi_{12,\mathrm{in}}}\bra{\Phi_{12,\mathrm{in}}}^{\otimes \kappa} = \left( \int d\mu(\psi_1) \ket{\psi_1}\bra{\psi_1}^{\otimes \kappa} \right) \times \left( \int d\mu(\psi_2) \ket{\psi_2}\bra{\psi_2}^{\otimes \kappa} \right),
\hat{\Omega}_{\overline{p}_0}^{(\kappa)} &=& \int d\mu(\psi_1, \psi_2) \ket{\Phi_{12,\mathrm{in}}}\bra{\Phi_{12,\mathrm{in}}}^{\otimes \kappa} \nonumber \\
    &=& \prod_{j=1}^2 \int d\mu(\psi_j) \ket{\psi_j}\bra{\psi_j}^{\otimes \kappa},
\label{eq:Omega2_p0}
\end{eqnarray}
Here, we can utilize the well-established Haar integral identities for pure states, which follow from the Schur–Weyl duality (or see {\bf Appendix~\ref{appendix:A}}): For a pure state $\ket{\psi} \in \mathcal{H}$ with $\dim(\mathcal{H}) = d$, the integration of the $\kappa$-fold powers of the pure states is given by
\begin{eqnarray}
\int d\mu(\psi) \ket{\psi}\bra{\psi}^{\otimes \kappa} = \frac{\hat{P}^{(d,\kappa)}_{\mathrm{sym}}}{\tr \hat{P}^{(d,\kappa)}_{\mathrm{sym}}},
\label{eq:Schur_iden}
\end{eqnarray}
where $\tr \hat{P}^{(d,\kappa)}_{\mathrm{sym}} = \binom{d+\kappa-1}{\kappa}$. Then, following the indexing rule for the subsystems described in Eq.~(\ref{eq:re_indices}) and using Eq.~(\ref{eq:Schur_iden}), 
\begin{eqnarray}
\int d\mu(\psi_1) \ket{\psi_1}\bra{\psi_1}^{\otimes 2} &=& \frac{2!}{d_1  (d_1 + 1)} \hat{P}_{13}^{+}, \nonumber \\
\int d\mu(\psi_2) \ket{\psi_2}\bra{\psi_2}^{\otimes 2} &=& \frac{2!}{d_2  (d_2 + 1)} \hat{P}_{24}^{+}.
\end{eqnarray}
A similar calculation can be performed for four copies.  Equipped with these,  we have 
\begin{eqnarray}
\hat{\Omega}_{\overline{p}_0}^{(2)} &=& \frac{(2!)^2 \hat{P}_{13}^{+} \hat{P}_{24}^{+}}{d_1 d_2 (d_1 + 1)(d_2 + 1)}, \nonumber \\
\hat{\Omega}_{\overline{p}_0}^{(4)} &=& \frac{(4!)^2 \hat{P}_{1357}^{+} \hat{P}_{2468}^{+}}{\prod_{i=1}^{2} d_i(d_i+1)(d_i+2)(d_i+3)}.
\end{eqnarray}
To proceed, we represent the linearized entropy and its square, such that
\begin{eqnarray}
E(\ket{\Psi_{12}}) &=& 2\tr \bigl( \ket{\Psi_{12}}\bra{\Psi_{12}}^{\otimes 2} \hat{P}_{13}^{-} \bigr), \nonumber \\
E^2(\ket{\Psi_{12}}) &=& 4\tr \bigl( \ket{\Psi_{12}}\bra{\Psi_{12}}^{\otimes 4} \hat{P}_{13}^{-} \hat{P}_{57}^{-} \bigr).
\end{eqnarray}
Then, by averaging $E(\ket{\Psi_{12}})$ over $\overline{p}_0$ via the product Haar measure $\mu(\psi_1,\psi_2) = \mu(\psi_1) \mu(\psi_2)$, 
\begin{widetext}
\begin{eqnarray}
e_{\bar{p}_0} (\hat{U}) &=& 2\expt{E( \ket{\Psi_{12}} )} \nonumber \\
    &=& 2 \int d \mu(\psi_1, \psi_2)  \tr \bigl( U^{\otimes 2}  \ket{\Phi_{12,\mathrm{in}}}\bra{\Phi_{12,\mathrm{in}}}^{\otimes 2} \hat{U}^{\dagger \otimes 2} \hat{P}_{13}^{-} \bigr),  \nonumber \\
    &=& 2 \tr \bigl( \hat{U}^{\otimes 2} \hat{\Omega}_{\bar{p}_0}^{(2)} \hat{U}^{\dagger \otimes 2} \hat{P}_{13}^{-} \bigr).
\end{eqnarray}
\end{widetext}
By performing an analogous calculation for $E(\ket{\Psi_{12}})^2$, we complete the proof. 

\end{proof}

{\em Remarks.}---{\bf Propoition~\ref{prop:1}} shows that for the uniform Haar distribution over product states, the second and fourth moment operators $\hat{\Omega}_{\overline{p}_0}^{(2)}$ and $\hat{\Omega}_{\overline{p}_0}^{(4)}$ are explicitly given by the projections onto the symmetric subspaces. Specifically, these moment operators reduce to the tensor products of the symmetric projectors [as in Eq.~(\ref{eq:Ometa_prop2})]:
\begin{eqnarray}
\hat{\Omega}_{\overline{p}_0}^{(2)} \propto \hat{P}_{13}^{+} \hat{P}_{24}^{+}~~\text{and}~~\hat{\Omega}_{\overline{p}_0}^{(4)} \propto \hat{P}_{1357}^{+} \hat{P}_{2468}^{+}.
\end{eqnarray}
This reflects the full permutation symmetry inherent in the Haar-random product states. This structure tells us that EP can be interpreted and estimated as the expectation of $\hat{U}^{\dagger \otimes 2} \hat{P}_{13}^{-} \hat{U}^{\otimes 2}$ over the second-moment symmetric states $\hat{\Omega}_{\overline{p}_0}^{(2)}$ and EPD can be viewed as a quantity associated with the expectation of $\hat{U}^{\dagger \otimes 4} \hat{P}_{13}^{-} \hat{P}_{57}^{-} \hat{U}^{\otimes 4}$ on the fourth moment state $\hat{\Omega}_{\overline{p}_0}^{(4)}$. These forms are particularly useful in analytical studies on the entanglement production behavior of a quantum operation, as they enable efficient computation by leveraging the algebraic properties of the permutation operators and projectors.

The explicit forms of EP and EPD derived in {\bf Proposition~\ref{prop:1}} allow us to precisely identify the structural conditions under which a unitary operation $\hat{U}$ fails to generate entanglement---either on average or in its fluctuations. In particular, they enable a direct link between the entangling capacity of $\hat{U}$ and its symmetry properties encoded via commutation relations with permutation operators. We formalize this connection in the following theorems:
\begin{theorem}
Let $\hat{U}$ be a nontrivial unitary operator acting on a bipartite Hilbert space. Then, the following relation holds:
\begin{eqnarray}
%e_{\overline{p}_0}(\hat{U})=0 \quad\text{if either}\quad \text{(i)}~\bigl[ \hat{U}^{\otimes 2}, \hat{P}_{13}^{+} \bigr] = 0 ~\text{or}~ \text{(ii)}~\bigl[ \hat{U}^{\otimes 2}, \hat{P}_{13}^{+}\hat{P}_{24}^{+} \bigr] = 0,
e_{\overline{p}_0}(\hat{U})=0
\end{eqnarray}
if either
\begin{eqnarray}
\mathrm{(i)}~\bigl[ \hat{U}^{\otimes 2}, \hat{P}_{13}^{+} \bigr] = 0 ~\text{or}~ \mathrm{(ii)}~\bigl[ \hat{U}^{\otimes 2}, \hat{P}_{13}^{+}\hat{P}_{24}^{+} \bigr] = 0,
\label{eq:thm1}
\end{eqnarray}
where $[ \hat{A}, \hat{B} ]  :=  \hat{A}\hat{B} - \hat{B}\hat{A}$ denotes the commutator.
\label{thm:1}
\end{theorem}

\begin{proof}---We begin by recalling the compact expression of EP derived from permutation operator formalism:
\begin{eqnarray}
e_{\overline{p}_0}(\hat{U}) = 2  (2!)^2  C_{d_1} C_{d_2} \tr \bigl( \hat{U}^{\otimes 2} \hat{P}_{13}^{+} \hat{P}_{24}^{+} \hat{U}^{\dagger \otimes 2} \hat{P}_{13}^{-} \bigr),
\label{eq:EP_re}
\end{eqnarray}
where the constants $C_{d_j}$ are as defined in Eq.~(\ref{eq:CDdj}). This expression results from averaging over Haar-random product states and encapsulates the entangling capability of $\hat{U}$ in terms of its action on symmetrized tensor spaces. Now, suppose condition (i) or (ii) in Eq.~(\ref{eq:thm1}) holds. Then, due to the commutation of $\hat{U}^{\otimes 2}$ with $\hat{P}_{13}^{+}$ or with $\hat{P}_{13}^{+}\hat{P}_{24}^{+}$, it follows that:
\begin{eqnarray}
\tr \bigl( \hat{U}^{\otimes 2} \hat{P}_{13}^{+} \hat{P}_{24}^{+} \hat{U}^{\dagger \otimes 2} \hat{P}_{13}^{-} \bigr) = \tr \bigl( \hat{P}_{13}^{+} \hat{P}_{24}^{+} \hat{P}_{13}^{-} \bigr) = 0,
\end{eqnarray}
where the vanishing trace arises from the orthogonality of the symmetric and antisymmetric projectors, i.e., $\hat{P}_{13}^{+} \hat{P}_{13}^{-} = 0$, as in Eq.~(\ref{eq:orth_P}). Therefore, $e_{\overline{p}_0}(\hat{U}) = 0$ directly follows from Eq.~(\ref{eq:EP_re}). 
\end{proof}

{\em Remarks.}---{\bf Theorem~\ref{thm:1}} provides a powerful and physically meaningful criterion: the vanishing of EP is dictated by the symmetry properties of $\hat{U}^{\otimes 2}$. Specifically, if $\hat{U}^{\otimes 2}$ commutes with either $\hat{P}_{13}^{+}$ or $\hat{P}_{13}^{+}\hat{P}_{24}^{+}$, then it fails to generate entanglement on average. Let us now unpack the implications of these commutation conditions:
\begin{itemize}
\item[(i)] Product unitaries: The condition (i), i.e., $\bigl[ \hat{U}^{\otimes 2}, \hat{P}_{13}^{+} \bigr] = 0$, is satisfied by any product unitary operation, i.e., $\hat{U} = \hat{u}_1 \otimes \hat{u}_2$ for local unitaries $\hat{u}_j$ ($j=1,2$) acting on subsystems $\mathcal{H}_j$ of input $\ket{\Psi}$. This result aligns perfectly with physical intuition: product unitaries cannot generate entanglement and hence must have zero EP.
\item[(ii)] Non-product yet non-entangling unitaries: Remarkably, condition (ii), i.e., $\bigl[ \hat{U}^{\otimes 2}, \hat{P}_{13}^{+} \hat{P}_{24}^{+} \bigr] = 0$, is strictly weaker than the condition (i)\footnote{The condition (i) implies the condition (ii) which comes from the commutator of operator $\hat{A}$ with the product of operators $\hat{B}$ and $\hat{C}$ can be expanded as follows $[\hat{A},\hat{B}\hat{C}] = [\hat{A},\hat{B}]\hat{C} + \hat{B}[\hat{A},\hat{C}]$.}. There exist non-product unitaries that violate (i) but still satisfy (ii), and thus yield $e_{\overline{p}_0}(\hat{U}) = 0$. A canonical example is the SWAP operation, which exchanges two subsystems but introduces no entanglement---despite being non-product in structure.
\end{itemize}
This distinction highlights a subtle yet critical point: not all non-product operations are entangling, and not all entanglement-free behavior can be attributed to the non-product alone. The commutation-based characterization in {\bf Theorem~\ref{thm:1}} offers useful framework to capture such nuanced structure, consistent with prior studies~\cite{PhysRevA.62.030301}.

%We emphasize that analyzing the local and symmetric invariance properties of $\hat{U}^{\otimes 2}$ provides a clear window into its entangling behavior. In this sense, the permutation-symmetry formalism not only yields computational tools but also unveils conceptual insights into when and why entanglement fails to emerge.

Then, we provide our second theorem as one of the main results:
\begin{theorem}
Given a non-zero unitary operator $\hat{U}$ acting on a bipartite system, the following equivalence holds:
\begin{eqnarray}
\Delta_{\overline{p}_0}(\hat{U})=0~\text{iff}~e_{\overline{p}_0}(\hat{U})=0.
\label{eq:thm2}
\end{eqnarray}
%$\Delta_{\overline{p}_0}(\hat{U})=0$ if $e_{\overline{p}_0}(\hat{U})=0$ and $\left[  \hat{U}^{\otimes 4}, \hat{P}^{+}_{1357} \right]=0$.
\label{thm:2}
\end{theorem}

\begin{proof}---We begin with the squared EPD in terms of the trace operator form, analogous to Eq.~(\ref{eq:EP_re}), as
\begin{widetext}
\begin{eqnarray}
\Delta_{\overline{p}_0}(\hat{U})^2  = 4 (4!)^2 D_{d_1} D_{d_2}  \tr \bigl( \hat{U}^{\otimes 4}  \hat{P}_{1357}^{+} \hat{P}_{2468}^{+} \hat{U}^{\dagger \otimes 4} \hat{P}_{13}^{-}  \hat{P}_{57}^{-} \bigr) - e_{\overline{p}_0}(\hat{U})^2 = \xi_1 \Gamma_1 - \xi_2 \Gamma_2,
\label{eq:EPD_re}
\end{eqnarray}
\end{widetext}
where we define
\begin{eqnarray}
\xi_1 &=& 4 (4!)^2 D_{d_1} D_{d_2}, \nonumber \\
\xi_2 &=& 4 (2!)^4 C_{d_1}^2 C_{d_2}^2, \nonumber \\
\Gamma_1 &=& \tr \bigl( \hat{U}^{\otimes 4}  \hat{P}_{1357}^{+} \hat{P}_{2468}^{+} \hat{U}^{\dagger \otimes 4} \hat{P}_{13}^{-}  \hat{P}_{57}^{-} \bigr) , \nonumber \\
\Gamma_2 &=& \tr \bigl( \hat{U}^{\otimes 4}  \hat{P}_{13}^{+} \hat{P}_{57}^{+} \hat{P}_{24}^{+} \hat{P}_{68}^{+} \hat{U}^{\dagger \otimes 4} \hat{P}_{13}^{-}  \hat{P}_{57}^{-} \bigr).
\label{eq:Gamma}
\end{eqnarray}
Note that $\xi_2 \Gamma_2 = e_{\overline{p}_0}(\hat{U})^2$ and in evaluating $\Gamma_2$, the following trick is used~\footnote{Note that $\tr\bigl(\hat{A} \otimes \hat{B}\bigr) = \bigl(\tr\hat{A}\bigr) \bigl(\tr\hat{B}\bigr)$ and $\hat{A}\hat{B} \otimes \hat{C}\hat{D} = \bigl( \hat{A} \otimes \hat{C} \bigr) \bigl( \hat{B} \otimes \hat{D} \bigr)$.}.
\begin{eqnarray}
&& \tr\bigl( \hat{U}^{\otimes 2} \hat{P}_{13}^{+} \hat{P}_{24}^{+} \hat{U}^{\dagger \otimes 2} \hat{P}_{13}^{-} \bigr)^2 \nonumber \\
&& \quad = \tr \bigl( \hat{U}^{\otimes 4}  \hat{P}_{13}^{+} \hat{P}_{57}^{+} \hat{P}_{24}^{+} \hat{P}_{68}^{+} \hat{U}^{\dagger \otimes 4} \hat{P}_{13}^{-}  \hat{P}_{57}^{-} \bigr).
\end{eqnarray}

Firstly, it is definitely true that if $e_{\overline{p}_0}(\hat{U})=0$, then $\Delta_{\overline{p}_0}(\hat{U})=0$. To establish the converse, suppose $e_{\overline{p}_0}(\hat{U}) \neq 0$. By construction, both $\xi_1$ and $\xi_2$ are strictly positive for $d_{1}, d_{2} \geq 1$, and satisfy $\xi_1 \geq \xi_2$ with equality only if $d_1 = d_2 = 1$. Furthermore, it is straightforward that the inequality $\Gamma_1 \geq \Gamma_2$ holds, with equality only if $\hat{U}=0$. Thus, for all nontrivial cases, i.e., $d_{1}, d_{2} > 1$ and any $\hat{U} \neq \text{Null}$, we have $\Delta_{\overline{p}_0}(\hat{U})^2 = \xi_1 \Gamma_1 - \xi_2 \Gamma_2 > 0$, implying $\Delta_{\overline{p}_0}(\hat{U}) \neq 0$. Consequently, if $\Delta_{\overline{p}_0}(\hat{U}) = 0$, it follows necessarily that $e_{\overline{p}_0}(\hat{U}) = 0$, completing the proof.
\end{proof}

From {\bf Theorem~\ref{thm:2}}, we see that the perfectly uniform entanglement generation (zero fluctuation over product inputs) is only possible when the average entangling power itself vanishes. A complementary and more quantitative interpretation of EPD is obtained by viewing it as a susceptibility of EP under a controlled biasing of the input ensemble.

\begin{corollary}[EPD as a susceptibility of entangling power]\label{corol:1}
For a fixed unitary operator $\hat{U}$, let $P_{\hat{U}}(e)$ denote the distribution of entanglement values $e:=E(\hat{U}\ket{\psi_1}\otimes\ket{\psi_2})$ induced by Haar-random product inputs (see Sec.~\ref{Sec:2} and Appendix~\ref{appendix:dist}). Define the moment-generating function
\begin{eqnarray}
Z_{\hat{U}}(\lambda) &:=& \int de P_{\hat{U}}(e)\, e^{\lambda e} \nonumber \\
&=& \int d\mu(\psi_1, \psi_2) e^{\lambda E(\hat{U}\ket{\psi_1}\otimes\ket{\psi_2})},
\label{eq:Z_lambda}
\end{eqnarray}
and the corresponding exponentially biased average (``biased entangling power'')
\begin{eqnarray}
e_{\overline{p}_0}^{(\lambda)}(\hat{U}) := \frac{\int de \,e P_{\hat{U}}(e) e^{\lambda e}}{\int de P_{\hat{U}}(e)\, e^{\lambda e}} = \frac{d}{d\lambda}\log Z_{\hat{U}}(\lambda).
\label{eq:ep_lambda}
\end{eqnarray}
Then, we attain
\begin{eqnarray}
\left.\frac{d}{d\lambda} e_{\overline{p}_0}^{(\lambda)}(\hat{U})\right|_{\lambda=0}=\Delta_{\overline{p}_0}(\hat{U})^2,
\label{eq:susceptibility}
\end{eqnarray}
where $e_{\overline{p}_0}^{(0)}(\hat{U})=e_{\overline{p}_0}(\hat{U})$.
\end{corollary}

\begin{proof}---By construction, $Z_{\hat{U}}(0)=1$ and Eq.~(\ref{eq:ep_lambda}) reduces to $e_{\overline{p}_0}^{(0)}(\hat{U})=\int de\, e\,P_{\hat{U}}(e)=e_{\overline{p}_0}(\hat{U})$.
Differentiating Eq.~(\ref{eq:ep_lambda}) once more and evaluating at $\lambda=0$, we obtain
\begin{eqnarray}
\left.\frac{d}{d\lambda} e_{\overline{p}_0}^{(\lambda)}(\hat{U})\right|_{\lambda=0} &=& \int de\, e^2 P_{\hat{U}}(e) - \left(\int de\, e P_{\hat{U}}(e)\right)^2 \nonumber \\
    &=& \Delta_{\overline{p}_0}(\hat{U})^2,
\end{eqnarray}
completing the proof.
\end{proof}

{\em Remarks.}---By synthesizing {\bf Theorems~\ref{thm:1}}, {\bf \ref{thm:2}}, and {\bf Corollary~\ref{corol:1}} presented above, we arrive at a fundamental principle: {\em it is inherently impossible for quantum operations to generate entanglement uniformly across all possible input states.} Indeed, this finding reveals a fundamental relation---any increase in entangling capability necessarily increases the dependence on, and bias toward, specific input states. This highlights a crucial limitation and a previously unexplored constraint within physics of the quantum entanglement generation. This insight significantly broadens our understanding of the entanglement, offering critical implications for quantum circuit design, optimization in quantum computing, and especially for quantum machine learning~\cite{Eisert2021entangling,Cerezo2022challenges,Azad2023qleet}.

Our analysis further emphasizes the physical significance of EPD alongside EP. By introducing and rigorously analyzing EPD, it is clearly established that the input-state dependence is an intrinsic factor in entanglement generation. Consequently, EPD emerges as an essential and fundamental metric for deepening our understanding of the quantum entanglement generation.

\subsection{A more closed-form of EP and EPD} %-------------------------------------------------------------------------------

We further formulate the expressions for EP and EPD into more computationally tractable forms. Our approach leverages the permutation symmetries, providing a unified framework for both EP and EPD. To start, we provide the following Lemma:
\begin{lemma}\label{lemma:1}
Let $A, B$ be the subsystem of total system $AB$ with the equal size, say, $\abs{A}=\abs{B}=\kappa$. Here, if $\alpha \in S_{A}$ and $\beta \in S_{B}$ have the same shape with respect to its relative positions, then the following holds:
\begin{eqnarray}
\bigl[ \hat{V}_d(\alpha) \hat{V}_d(\beta), \hat{U}^{\otimes \kappa} \bigr]=0
\end{eqnarray}
where $\hat{V}_d$ is the permutation matrices of total system, introduced in {\bf Definition~\ref{def:1}}.
\end{lemma}

\begin{proof}---In the context of our analysis, we rewrite $\hat{U}$ explicitly as $\hat{U}_{12}$. Then, the $\kappa$-fold tensor product of $\hat{U}$ can be consistently represented as
\begin{eqnarray}
\hat{U}_{12}^{\otimes \kappa} = \hat{U}_{12} \otimes \hat{U}_{34} \otimes \cdots \otimes \hat{U}_{2\kappa-1, 2\kappa},
\end{eqnarray}
with 
\begin{eqnarray}
\hat{U}_{12} = \hat{U}_{34} = \cdots = \hat{U}_{2\kappa-1, 2\kappa}.
\end{eqnarray}
Given that the total system is indexed by the set $\{ 1,2,\dots,2\kappa \}$, we identify subsystem $A$ as the odd-index subset $\{1,3,\dots,2\kappa-1\}$, and subsystem $B$ as the even-index subset $\{2,4,\dots,2\kappa\}$. Here, we let $\alpha \in S_A$ and $\beta \in S_B$ denote disjoint cycles within $S_{2\kappa}$. Then, utilizing $\bigl[ \hat{V}_d(\alpha), \hat{V}_d(\beta) \bigr]=0$ and Eq.~(\ref{eq:conjVdA2}), we complete the proof.
\end{proof}

Leveraging the commutation relations established in {\bf Lemma~\ref{lemma:1}}, we simplify the expression for $e_{\overline{p}_0}(\hat{U})$ to obtain the following proposition~\cite{PhysRevA.62.030301,PhysRevA.63.040304}:
\begin{proposition}\label{prop:2}
Given a non-zero unitary operator $\hat{U}$ acting on a bipartite system and considering a uniform distribution $\overline{p}_0$ over the input product states, EP can be expressed as
\begin{eqnarray}
e_{\overline{p}_0}(\hat{U})  = \frac{d_1 d_2 \left( E(\hat{U}) + \tilde{E}(\hat{U}) + \frac{1}{d_1 d_2} - 1 \right) }{\left( d_1 + 1 \right) \left( d_2 + 1 \right)} ,
\label{eq:w1}
\end{eqnarray}
where  $E$ and $\tilde{E}$ are the linear operator entanglement entropies, defined by
\begin{eqnarray}
E(\hat{U}) &=& 1 - \frac{1}{d_1^2 d_2^2} \tr \bigl( \hat{U}^{\otimes 2} \hat{P}_{13} \hat{U}^{\dagger \otimes 2} \hat{P}_{13} \bigr), \nonumber \\
\tilde{E}(\hat{U}) &=& 1 - \frac{1}{d_1^2 d_2^2} \tr \bigl( \hat{U}^{\otimes 2} \hat{P}_{24} \hat{U}^{\dagger \otimes 2} \hat{P}_{13} \bigr).
\label{eq:w2}
\end{eqnarray}
In particular, for the symmetric case $d = d_1 = d_2$,
\begin{eqnarray}
e_{\overline{p}_0}(\hat{U}) = \frac{d^2}{(d+1)^2} \left( E(\hat{U}) + E(\hat{U}\hat{S}) - E(\hat{S}) \right),
\label{eq:ep0inE}
\end{eqnarray}
where $\hat{S} = \hat{P}_{12}$ is the SWAP operator, and the entropies are explicitly given by
\begin{eqnarray}
E(U) &=& 1-\frac{1}{d^4} \tr \bigl( \hat{U}^{\otimes 2} \hat{P}_{13} \hat{U}^{\dagger \otimes 2} \hat{P}_{13} \bigr), \nonumber \\
E(\hat{U}\hat{S}) &=& \tilde{E}(\hat{U}) = 1 - \frac{1}{d^4} \tr \bigl( \hat{U}^{\otimes 2} \hat{P}_{24} \hat{U}^{\dagger \otimes 2} \hat{P}_{13} \bigr).
\label{eq:w4}
\end{eqnarray}
\end{proposition}

\begin{proof}---From {\bf Lemma~\ref{lemma:1}}, we obtain the relation
\begin{widetext}
\begin{eqnarray}
4 \tr \bigl( \hat{U}^{\otimes 2} \hat{P}_{13}^{+} \hat{P}_{24}^{+} \hat{U}^{\dagger \otimes 2} \hat{P}_{13} \bigr) = \tr \bigl( \hat{P}_{13} + \hat{P}_{24} + \hat{U}^{\otimes 2} \hat{P}_{13} \hat{U}^{\dagger \otimes 2} \hat{P}_{13} + \hat{U}^{\otimes 2} \hat{P}_{24} U^{\dagger \otimes 2} \hat{P}_{13} \bigr).
\label{eq:keyforepinE}
\end{eqnarray}
\end{widetext}
Utilizing the linearity of the trace, along with the identities $\tr \hat{P}_{13} = d_1 d_2^2$ and $\tr \hat{P}_{24} = d_1^2 d_2$, we can directly verify Eqs.~(\ref{eq:w1}) and (\ref{eq:w2}). For the case $d = d_1 = d_2$, using the property $\hat{S}^{\otimes 2} \hat{P}_{13} \hat{S}^{\dagger \otimes 2} = \hat{P}_{24}$, we verify Eq.(\ref{eq:w4}) and explicitly compute
\begin{eqnarray}
E(S) = 1 - \frac{1}{d^4} \tr \bigl( \hat{S}^{\otimes 2} \hat{P}_{13} \hat{S}^{\dagger \otimes 2} \hat{P}_{13} \bigr) = 1 - \frac{1}{d^2}.
%   &=& 1 - \frac{1}{d^4} \tr \bigl( \hat{P}_{24} \hat{P}_{13} \bigr) = 1 - \frac{1}{d^2}.
\end{eqnarray}
Combining all these results, we finally obtain Eq.~(\ref{eq:ep0inE}), completing the proof.
\end{proof}

Building upon the closed-form expression obtained previously for EP, we now focus on EPD. EPD can be represented by a structurally analogous expression involving permutation operators acting on higher-order tensor products, as stated in the following proposition:
\begin{widetext}
\begin{proposition}\label{prop:3}
Given a non-zero unitary operator $\hat{U}$ acting on a bipartite system and assuming a uniform distribution $\overline{p}_0$ over the input product states, EPD is given by
\begin{eqnarray}
\Delta_{\overline{p}_0}(\hat{U}) = \sqrt{ D_{d_1} D_{d_2} \left( F_{id} - 2 F_{(13)} + F_{(13)(57)} \right)  - e_{\overline{p}_0}(\hat{U})^2},
\end{eqnarray}
where $id$, $(13)$, and $(13)(57)$ denote cycles in $S_8$, and
\begin{eqnarray}
F_{\pi} = \sum_{\nu_1 \in S_{A}} \sum_{\nu_2 \in S_B} \tr \bigl[  \bigl( \hat{U}^{\otimes 4} \hat{U}^{\dagger \otimes 4} \bigr)_{(\nu_1 \circ \nu_2)^{-1}} \hat{P}_{\nu_1 \circ \nu_2 \circ \pi} \bigr],
\label{eq:F_pi}
\end{eqnarray}
Here, $A$ and $B$ are the subspaces identified by the odd-index subset $\{1,3,5,7\}$ and the even-index subset $\{2,4,6,8\}$, respectively, and $\pi \in S_{8}$.
\end{proposition}
\end{widetext}

\begin{proof}---Recalling the expression for the 4-copy case, we have for subsystems $A$ and $B$:
\begin{eqnarray}
&& 4 \tr \bigl( \hat{U}^{\otimes 4} \Omega_{\overline{p}_0}^{(4)} \hat{U}^{\dagger \otimes 4} \hat{P}_{13}^{-}  \hat{P}_{57}^{-} \bigr) \nonumber \\
&& \quad = D_{d_1} D_{d_2} \sum_{\nu_1 \in S_A} \sum_{\nu_2 \in S_B} \left( \Xi_1 - \Xi_2 - \Xi_3 + \Xi_4 \right),
\end{eqnarray}
where the factors $\Xi_j$ ($j=1,2,3,4$) are defined as
\begin{eqnarray}
\Xi_1 &=& \tr \bigl( \hat{U}^{\otimes 4} \hat{P}_{\nu_1} \hat{P}_{\nu_2} \hat{U}^{\dagger \otimes 4} \bigr), \nonumber \\
\Xi_2 &=& \tr \bigl( \hat{U}^{\otimes 4} \hat{P}_{\nu_1} \hat{P}_{\nu_2} \hat{U}^{\dagger \otimes 4} \hat{P}_{13} \bigr), \nonumber \\
\Xi_3 &=& \tr \bigl( \hat{U}^{\otimes 4} \hat{P}_{\nu_1} \hat{P}_{\nu_2} \hat{U}^{\dagger \otimes 4} \hat{P}_{57} \bigr), \nonumber \\
\Xi_4 &=& \tr \bigl( \hat{U}^{\otimes 4} \hat{P}_{\nu_1} \hat{P}_{\nu_2} \hat{U}^{\dagger \otimes 4} \hat{P}_{13} \hat{P}_{57} \bigr).
\end{eqnarray}
By renaming indices and exploiting permutation symmetry, we verify that
\begin{eqnarray}
&& \sum_{\nu_1 \in S_A} \sum_{\nu_2 \in S_B}  \tr \bigl( \hat{U}^{\otimes 4} \hat{P}_{\nu_1} \hat{P}_{\nu_2} \hat{U}^{\dagger \otimes 4} \hat{P}_{13} \bigr) \nonumber \\
&& \quad = \sum_{\nu_1 \in S_A} \sum_{\nu_2 \in S_B}  \tr \bigl( \hat{U}^{\otimes 4} \hat{P}_{\nu_1} \hat{P}_{\nu_2} \hat{U}^{\dagger \otimes 4} \hat{P}_{57} \bigr).
\end{eqnarray}
Furthermore, for $\nu_1 \in S_A$, $\nu_2 \in S_B$, and $\pi \in S_8$, we define
\begin{eqnarray}
F_{\pi} &=& \tr \bigl( \hat{U}^{\otimes 4} \hat{P}_{\nu_1} \hat{P}_{\nu_2} \hat{U}^{\dagger \otimes 4} \hat{P}_{\pi} \bigr) \nonumber \\
     &=& \tr \bigl[ \bigl( \hat{U}^{\otimes 4} \hat{U}^{\dagger \otimes 4} \bigr)_{(\nu_1 \circ \nu_2)^{-1}}  \hat{P}_{\nu_1 \circ \nu_2 \circ \pi} \bigr].
\end{eqnarray}
In deriving this, it is utilized: (i) reordering of subsystems by $\hat{V}_d(\pi)$ as in Eq.~(\ref{eq:conjVdA}), (ii) the fact that since $S_A$ and $S_B$ are subgroups of $S_8$, their elements $\nu_1, \nu_2$ belong to $S_8$, ensuring well-defined compositions, and (iii) commutativity of operators corresponding to disjoint permutations $\nu_1, \nu_2$, i.e., $\bigl[ \hat{P}_{\nu_1}, \hat{P}_{\nu_2} \bigr] = 0$. Combining these points, we obtain the desired compact expression for $\Delta_{\overline{p}_0}(\hat{U})$ in terms of the cycle permutations in $S_8$, as claimed.
\end{proof}

%-------------------------------------------------------------------------------------------------------------------------------------------------------------------------------------------------------------------------------------
\section{Entangling Power and Entangling Power Deviation of Two-Qubit Gates}\label{Sec:4}
%-------------------------------------------------------------------------------------------------------------------------------------------------------------------------------------------------------------------------------------

Two-qubit gates combined with single-qubit operations constitute the fundamental building blocks for universal quantum computation. Thus, accurately characterizing their entangling properties is essential for both theoretical insights and practical implementations in quantum computation. While extensive studies have only focused on EP of the two-qubit gates \cite{PhysRevA.110.062422, PhysRevResearch.2.043126,PhysRevA.83.062320,1204.0996,PhysRevA.78.052305,Shen_2018}, the input-state dependence, or a uniformity, in entanglement generation---referred to here as EPD---remain unexplored. Thus, in this section, we apply the analytical framework developed in the previous section to explicitly compute both EP and EPD for several representative classes of two-qubit gates. The chosen gate families reflect distinct physical interaction:
\begin{itemize}
    \item The controlled-unitary (CU) gate family designed using the Ising interaction~\cite{Debnath2016demonstration}, including the special cases, such as the controlled-NOT (CNOT) and controlled-Phase (CP) gates.
    \item The $\textrm{SWAP}^{\alpha}$ gate family derived from the (isotropic) exchange interaction~\cite{PhysRevA.78.052305,PhysRevA.72.052323}, which interpolates between the identity and SWAP---also referred to as the Power-of-SWAP gates.
    \item The $\textrm{iSWAP}$ gate family based on the ($XX \pm YY$)-type exchange interactions~\cite{PhysRevResearch.2.033447,PhysRevA.96.062323}, includes an additional relative-phase parameter. It still can realize all gates within the SWAP lines.
    \item The general two-qubit entangling gates from SU($4$), capturing the full nonlocal characteristics of arbitrary two-qubit unitaries~\cite{PhysRevResearch.2.043126,PhysRevA.70.052313, Shen_2018,PhysRevA.63.062309, PhysRevA.67.042313}.
\end{itemize}
These examples clearly illustrate the broad applicability of our formalism in systematically characterizing the entangling capabilities of quantum gates across diverse interaction types and operational contexts.

\subsection{Controlled-unitary (CU) gate family}  %-------------------------------------------------------------------------------

We examine the family of CU gates, where an arbitrary single-qubit unitary $\hat{u}$ is conditionally applied to the target qubit if the control qubit is in state $\ket{1}$, and the identity otherwise:
\begin{eqnarray}
\hat{U}_{CU} = \ket{0}\bra{0} \otimes \hat{\openone} + \ket{1}\bra{1} \otimes \hat{u}.
\end{eqnarray}
The CU gates is a fundamental component of quantum computing and appear widely in quantum algorithms. Analyzing their EP and EPD reveals how asymmetry and control-based interactions shape the distribution of entanglement generation. Specific choices of $\hat{u}$ recover well-known two-qubit gates such as CNOT and CP.

%------------
{\em Controlled-NOT (CNOT) Gate.}---The CNOT gate corresponds to the case $\hat{u} = \hat{\sigma}_x$, where $\hat{\sigma}_x$ is the Pauli-$X$ operator. Its matrix representation is
\begin{eqnarray}
U_\text{CNOT} = \left(
\begin{array}{cccc}
1 & 0 & 0 & 0 \\
0 & 1 & 0 & 0 \\
0 & 0 & 0 & 1 \\
0 & 0 & 1 & 0
\end{array}
\right).
\end{eqnarray}
Using {\bf Proposition~\ref{prop:2}} and {\bf Proposition~\ref{prop:3}}, EP and EPD of the CNOT gate are computed as
\begin{eqnarray}
e_{\overline{p}_0}(\hat{U}_\text{CNOT}) &=& \frac{2}{9}, \nonumber \\
\Delta_{\overline{p}_0}(\hat{U}_\text{CNOT}) &=& \frac{2\sqrt{11}}{45},
\label{eq:EPD_CNOT}
\end{eqnarray}
where the EP attains the maximum value for any two-qubit gate, as argued in prior works~\cite{PhysRevA.62.030301}.

%------------
{\em Controlled-phase (CP) gate.}---The CP gate corresponds to $\hat{u} = \mathrm{diag}(1, e^{i\theta})$ with $\theta \in [0, 2\pi)$, and its matrix form is\begin{eqnarray}
\hat{U}_\text{CP} = \left(
\begin{array}{cccc}
1 & 0 & 0 & 0 \\
0 & 1 & 0 & 0 \\
0 & 0 & 1 & 0 \\
0 & 0 & 0 & e^{i \theta}
\end{array}
\right).
\end{eqnarray}
The corresponding EP and EPD are given by
\begin{eqnarray}
e_{\overline{p}_0}(\hat{U}_\text{CP}) &=& \frac{2}{9} \sin^2\frac{\theta}{2}, \nonumber \\
\Delta_{\overline{p}_0}(\hat{U}_\text{CP}) &=& \frac{2\sqrt{11}}{45} \sin^2\frac{\theta}{2},
\label{eq:EPD_CP}
\end{eqnarray}
which reach their maxima when $\sin^2\frac{\theta}{2} = 1$, i.e., when $\hat{u} = \hat{\sigma}_z$. In this case, the EP and EPD values match those of the CNOT gate [see Eq.(\ref{eq:EPD_CNOT})], consistent with the previous studies on EP of two-qubit gates, while the EPD analysis can offer additional insights.

%------------
{\em General control-unitary (CU) gate.}---We now consider a general control-unitary gate, where the target unitary $\hat{u} \in \mathrm{SU}(2)$ is parameterized as
\begin{eqnarray}
\hat{u} = \left(
\begin{array}{cc}
e^{i(\delta + \frac{\alpha}{2} + \frac{\beta}{2})} \cos\frac{\theta}{2} & e^{i(\delta + \frac{\alpha}{2} - \frac{\beta}{2})} \sin\frac{\theta}{2} \\
-e^{i(\delta - \frac{\alpha}{2} + \frac{\beta}{2})} \sin\frac{\theta}{2} & e^{i(\delta - \frac{\alpha}{2} - \frac{\beta}{2})} \cos\frac{\theta}{2}
\end{array}
\right),
\end{eqnarray}
where $\delta$ is a global phase (physically irrelevant but retained for completeness), $\alpha$ and $\beta$ are $Z$-rotation angles before and after the $Y$-rotation, and $\theta$ is the rotation angle on the Bloch sphere. The corresponding controlled-unitary (CU) gate is then
\begin{eqnarray}
\hat{U}_\text{CU} \!=\! \left(
\begin{array}{cccc}
1 & 0 & 0 & 0 \\
0 & 1 & 0 & 0 \\
0 & 0 & e^{i(\delta + \frac{\alpha}{2} + \frac{\beta}{2})} \cos\frac{\theta}{2} & e^{i(\delta + \frac{\alpha}{2} - \frac{\beta}{2})} \sin\frac{\theta}{2} \\
0 & 0 & -e^{i(\delta - \frac{\alpha}{2} + \frac{\beta}{2})} \sin\frac{\theta}{2} & e^{i(\delta - \frac{\alpha}{2} - \frac{\beta}{2})} \cos\frac{\theta}{2}
\end{array}
\right).
\end{eqnarray}
\begin{widetext}
Its EP and EPD are given by
\begin{eqnarray}
e_{\overline{p}_0}(\hat{U}_\text{CU}) &=& \frac{5}{9} - \frac{1}{9} \left( 3 + \cos^2 \frac{\theta}{2} \left( 1 + \cos(\alpha + \beta) \right) \right), \nonumber \\
\Delta_{\overline{p}_0}(\hat{U}_\text{CU}) &=& \frac{\sqrt{11}}{9} - \frac{\sqrt{11}}{45} \left( 3 + \cos^2 \frac{\theta}{2} \left( 1 + \cos(\alpha + \beta) \right) \right).
\end{eqnarray}
Here, the extremal values are attained as follows:
\begin{eqnarray}
\cos^2\frac{\theta}{2} \left( 1 + \cos(\alpha + \beta) \right) =
\left\{ 
\begin{array}{lll}
0 & \Rightarrow & e_{\overline{p}_0}^\text{max}(\hat{U}_\text{CU}) = \frac{2}{9},~~\Delta_{\overline{p}_0}^\text{max}(\hat{U}_\text{CU}) = \frac{2\sqrt{11}}{45}, \\
\\
2 & \Rightarrow & e_{\overline{p}_0}^\text{min}(\hat{U}_\text{CU}) = \Delta_{\overline{p}_0}^\text{min}(\hat{U}_\text{CU}) = 0.
\end{array}
\right.
\end{eqnarray}
\end{widetext}
Remarkably, the ratio between EPD and EP, defined as 
\begin{eqnarray}
\eta:=\frac{\Delta_{\overline{p}_0}(\hat{U})}{e_{\overline{p}_0}(\hat{U})},
\end{eqnarray}
remains invariant across the entire CU gate family:
\begin{eqnarray}
%\eta_\text{CU} &=& \frac{\Delta_{\overline{p}_0}(\hat{U}_\text{CU})}{e_{\overline{p}_0}(\hat{U}_\text{CU})} = \frac{\Delta_{\overline{p}_0}(\hat{U}_\text{CNOT})}{e_{\overline{p}_0}(\hat{U}_\text{CNOT})} = \frac{\Delta_{\overline{p}_0}(\hat{U}_\text{CP})}{e_{\overline{p}_0}(\hat{U}_\text{CP})} \nonumber \\
\eta_\text{CU} = \frac{\Delta_{\overline{p}_0}(\hat{U}_\text{CU})}{e_{\overline{p}_0}(\hat{U}_\text{CU})} = \frac{\sqrt{11}}{5}.
\end{eqnarray}
This constancy of $\eta_\text{CU}$ reveals a linear EP--EPD law underpinning all CU-class gates: as the average entangling strength increases, the input-state variability increases proportionally. This provides a closed-form example of the intrinsic link between EP and EPD ({\bf Theorem~\ref{thm:2}} and the susceptibility interpretation in {\bf Corollary~\ref{corol:1}}.

\subsection{$\mathrm{SWAP}^\alpha$ gate family}   %-------------------------------------------------------------------------------

We analyze the entangling properties of the $\mathrm{SWAP}^\alpha$ gate family, which smoothly interpolates between the identity ($\alpha = 0$) and the full SWAP ($\alpha = 1$) gates via a continuous parameter $\alpha \in [0,1]$. The unitary matrix representation is given by
\begin{eqnarray}
\hat{U}_{\text{SWAP}^\alpha} =
\left(
\begin{array}{cccc}
1 & 0 & 0 & 0 \\
0 & \frac{1 + e^{i\pi\alpha}}{2} & \frac{1 - e^{i\pi\alpha}}{2} & 0 \\
0 & \frac{1 - e^{i\pi\alpha}}{2} & \frac{1 + e^{i\pi\alpha}}{2} & 0 \\
0 & 0 & 0 & 1
\end{array}
\right).
\end{eqnarray}
Then, from {\bf Proposition~\ref{prop:2}} and {\bf Proposition~\ref{prop:3}}, the corresponding EP and EPD for $U_{\mathrm{SWAP}^\alpha}$ are
\begin{eqnarray}
e_{\overline{p}_0}(\hat{U}_{\text{SWAP}^\alpha}) &=& \frac{1}{6} \sin^2(\pi \alpha), \nonumber \\
\Delta_{\overline{p}_0}(\hat{U}_{\text{SWAP}^\alpha}) &=& \frac{\sqrt{5}}{15} \sin^2(\pi \alpha).
\end{eqnarray}
Both quantities increase monotonically for $\alpha \in [0, \frac{1}{2}]$, reaching their maximum value at $\alpha = \frac{1}{2}$ (i.e., the $\sqrt{\mathrm{SWAP}}$ gate), and then symmetrically decrease toward zero at $\alpha = 0$ and $\alpha = 1$, corresponding to the identity and SWAP gates, respectively. Here, the maximum values are
\begin{eqnarray}
e_{\overline{p}_0}^\text{max}(\hat{U}_{\text{SWAP}^\alpha}) &=& \frac{1}{6}, \nonumber \\
\Delta_{\overline{p}_0}^\text{max}(\hat{U}_{\text{SWAP}^\alpha}) &=& \frac{\sqrt{5}}{15}.
\end{eqnarray}

While these EP results are consistent with the prior studies~\cite{PhysRevA.78.052305}, the analysis of EPD reveals additional insight. Remarkably, as with the CU class, the EP-to-EPD ratio remains constant throughout the $\mathrm{SWAP}^\alpha$ family:
\begin{equation}
\eta_{\text{SWAP}^\alpha}=\frac{\Delta_{\overline{p}_0}(\hat{U}_{\text{SWAP}^\alpha})}{e_{\overline{p}_0}(\hat{U}_{\text{SWAP}^\alpha})} = \frac{2\sqrt{5}}{5}.
\end{equation}
Similar to the CU gate class, we attain the constant EP-to-EPD ratio in the $\mathrm{SWAP}^\alpha$ family. Notably, the value $\eta{\text{SWAP}^\alpha} = \frac{2\sqrt{5}}{5}$ exceeds that of the CU family, $\eta_{\text{CU}} = \frac{\sqrt{11}}{5}$, indicating a stronger sensitivity to input states in the $\mathrm{SWAP}^\alpha$ gates. This contrast highlights a fundamental difference in how these gate families generate the entanglement---a point we explore further in the subsequent analyses.

\subsection{$\mathrm{iSWAP}$ gate family}   %-------------------------------------------------------------------------------

Here we consider another generalized two-qubit gate family, known as the $\textrm{iSWAP}$ gates,
\begin{eqnarray}
\hat{U}_{\text{iSWAP}}(\theta, \varphi) =
\begin{pmatrix}
1 & 0 & 0 & 0 \\
0 & \cos\frac{\theta}{2} & i e^{i\varphi} \sin\frac{\theta}{2} & 0 \\
0 & i e^{-i\varphi} \sin\frac{\theta}{2} & \cos\frac{\theta}{2} & 0 \\
0 & 0 & 0 & 1
\end{pmatrix},
\end{eqnarray}
where $\theta \in [0, \pi]$ controls the mixing the strength between $\ket{01}$ and $\ket{10}$, and $\varphi$ encodes an additional phase. Note that, for $\theta = \pi$ and $\varphi = \frac{3\pi}{2} + 2\pi n$, this gate reduces to the standard SWAP operation. Then, EP and EPD of this family are given by
%\begin{widetext}
\begin{eqnarray}
e_{\overline{p}_0}(\hat{U}_\text{iSWAP}) \!\!&=&\!\! \frac{2}{9} \sin^2\frac{\theta}{2} \left( 2 - \sin^2\frac{\theta}{2} \right), \nonumber \\
\Delta_{\overline{p}_0}(\hat{U}_\text{iSWAP}) \!\!&=&\!\! \frac{2}{45} \sin^2\frac{\theta}{2} \sqrt{34 + 30 \cos\theta + 7 \cos(2\theta)}. \nonumber \\
\end{eqnarray}
%\end{widetext}
Both quantities are independent of $\varphi$, and attain their maximum at $\theta = \pi$, yielding
\begin{eqnarray}
e_{\overline{p}_0}^\text{max}(\hat{U}_\text{iSWAP}) &=& \frac{2}{9}, \nonumber \\
\Delta_{\overline{p}_0}^\text{max}(\hat{U}_\text{iSWAP}) &=& \frac{2\sqrt{11}}{45},
\end{eqnarray}
which coincide with those of the CU class. However, unlike CU or $\mathrm{SWAP}^\alpha$ families, the EP-to-EPD ratio here, i.e., $\eta_{\text{iSWAP}} = \frac{\Delta_{\overline{p}_0}(\hat{U}_\text{iSWAP})}{e_{\overline{p}_0}(\hat{U}_\text{iSWAP})}$, is no longer constant and instead varies nonlinearly with the rotation angle $\theta$. This deviation suggests a more intricate relationship between the amount and uniformity of the generated entanglement by this gate family.

\subsection{Remarks: Distinct entangling profiles of $\mathrm{CU}$, $\mathrm{SWAP}^{\alpha}$, and $\mathrm{iSWAP}$ gates}  %-------------------------------------------------------------------------------

 \begin{figure}[t]  
 \centering
 \includegraphics[width=0.46\textwidth]{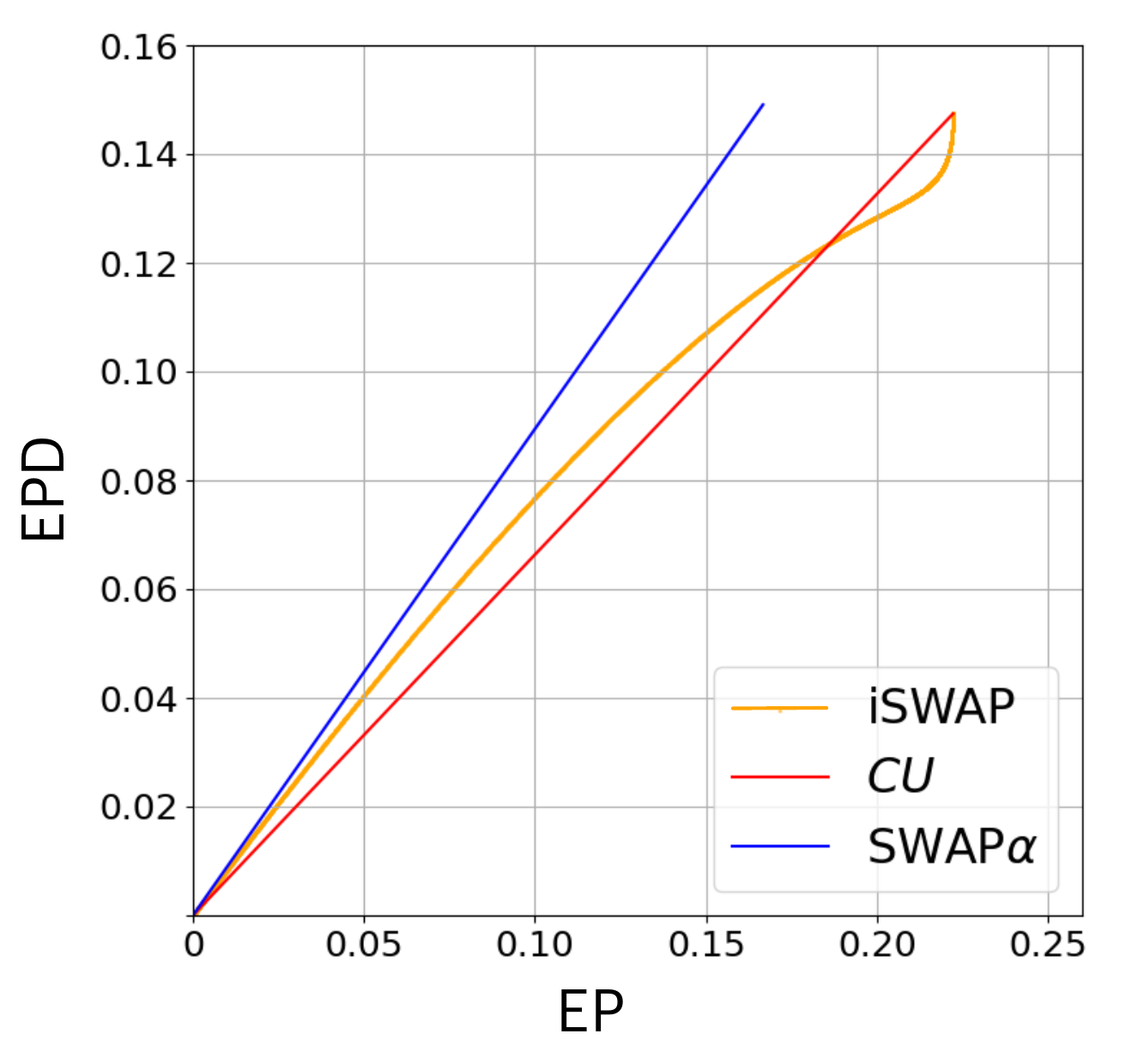}
\caption{EP vs. EPD graphs for physically implementable two-qubit gates. EP and EPD graphs are depicted for $\mathrm{iSWAP}$, CU, and $\mathrm{SWAP}^{\alpha}$ gates with respect to the realistic Hamiltonian parameters. CU and $\mathrm{SWAP}^{\alpha}$ exhibit linear EP--EPD relations, but with distinct profiles: $\mathrm{SWAP}^{\alpha}$ shows symmetric variation, while CU increases monotonically. Their EP-to-EPD ratios and maxima differ. In contrast, $\mathrm{iSWAP}$ displays a nonlinear profile, yet reaches the same maximal EP and EPD as CU. See Tab.~\ref{tab:1} for EP-to-EPD ratios.}
 \label{fig:EP-EPD_2q}
\end{figure}

\begin{table*}
\centering
\setlength{\tabcolsep}{0.20in}
\renewcommand{\arraystretch}{2.2}
\begin{tabular}{c  c  c}
\hline\hline
Class					 		& EP-to-EPD ratio $\eta:={\Delta_{\overline{p}_0}(\hat{U})}/{e_{\overline{p}_0}(\hat{U})}$ 		& ($\mathrm{EP}^\text{max}$, $\mathrm{EPD}^\text{max}$)	\\
\hline
$\mathrm{CU}$ (CNOT, CP, etc) 		& $\eta_\text{CU}= \frac{\sqrt{11}}{5}$								  			& $\left( \frac{2}{9}, \frac{2\sqrt{11}}{45} \right)$				\\
$\mathrm{SWAP}^\alpha$				& $\eta_{\text{SWAP}^\alpha}=\frac{2\sqrt{5}}{5}$  									& $\left( \frac{1}{6}, \frac{\sqrt{5}}{15} \right)$				\\
$\mathrm{iSWAP}(\theta, \varphi)$		& $\eta_\text{iSWAP}=\frac{2\sqrt{34 + 30\cos\theta + 7\cos(2\theta)}}{5(3 + \cos\theta)}$  	& $\left( \frac{2}{9}, \frac{2\sqrt{11}}{45} \right)$				\\
\hline\hline
\end{tabular}
\caption{The EP-to-EPD ratios and maxima for CU, $\mathrm{SWAP}^\alpha$, and $\mathrm{iSWAP}$ gate families. CU and $\mathrm{SWAP}^\alpha$ exhibit constant EP-to-EPD ratios due to their linear EP--EPD relations, whereas $\mathrm{iSWAP}$ shows a nonlinear profile. Notably, CU and $\mathrm{iSWAP}$ have the same maxima values (see main text for details).}
\label{tab:1}
\end{table*}

The $\mathrm{CU}$, $\mathrm{SWAP}^\alpha$, and $\mathrm{iSWAP}$ gate families exhibit qualitatively distinct entanglement-generation behaviors. While both $\mathrm{CU}$ and $\mathrm{SWAP}^\alpha$ gates show a linear relationship between EP and EPD, the $\mathrm{iSWAP}$ family displays a nonlinear and more intricate EP--EPD profile. In particular, $\mathrm{SWAP}^\alpha$ gates exhibit symmetric increase and decrease in EP and EPD, whereas $\mathrm{CU}$ gates show a proportional increase toward their maxima. Notably, the EP-to-EPD ratio differs between $\mathrm{CU}$ and $\mathrm{SWAP}^\alpha$, and their maximum values of EP and EPD are also distinct. Interestingly, $\mathrm{CU}$ and $\mathrm{iSWAP}$ gates, despite their fundamentally different structures, attain the same maximal EP and EPD. These comparisons are quantitatively summarized in Tab~\ref{tab:1} and Fig~\ref{fig:EP-EPD_2q}.

These differences are not merely mathematical but reflect the underlying physics by which each gate generates the entanglement. The strength and input-state sensitivity (or uniformity) of entanglement are closely tied to the physical interactions used for its implementation, i.e., the interaction Hamiltonian. Most importantly, such the distinctions are invisible when considering EP alone, but become apparent through the lens of EPD. This is because while EP measures the quantity of entanglement, EPD can capture its structural quality---i.e., how uniformly the entanglement is distributed across input states. Thus, the introduction of EPD is not a mere technical addition as a secondary measure, but a conceptually meaningful extension that enables a deeper, more refined classification of the entangling behaviors.

\subsection{Global EP and EPD structure of two-qubit gates in SU($4$)}  %-------------------------------------------------------------------------------

Any two-qubit unitary operation belongs to the Lie group SU($4$), the set of $4 \times 4$ special unitary matrices with determinant one. To systematically analyze and classify their entanglement-generating properties, it is fruitful to employ the so-called KAK decomposition~\cite{helgason1979}---a structural theorem for semi-simple Lie groups such as SU($4$). This decomposition allows any $\hat{U} \in \mathrm{SU}(4)$ to be expressed as a product of single-qubit local unitaries and a canonical nonlocal operation:
\begin{eqnarray}
%\hat{U} = \left( \hat{A}_2 \otimes \hat{B}_2 \right) \exp\left( - i \sum_{k=1}^{3} c_k \left( \hat{\sigma}_k \otimes \hat{\sigma}_k \right) \right) \left( \hat{A}_1 \otimes \hat{B}_1 \right)
\hat{U} = \left( \hat{A}_2 \otimes \hat{B}_2 \right) \hat{d}_\chi \left( \hat{A}_1 \otimes \hat{B}_1 \right)
\end{eqnarray}
where $\hat{A}_j$ and $\hat{B}_j$ ($j = 1, 2$) are single-qubit unitaries, and $\hat{d}_\chi$ captures all nonlocal (i.e., entangling) properties. The operation $\hat{d}_\chi$ is explicitly given by
\begin{widetext}
\begin{eqnarray}
\hat{d}_\chi = \exp\left( - i \sum_{k=1}^{3} \beta_k \left( \hat{\sigma}_k \otimes \hat{\sigma}_k \right) \right) 
= \left(
\begin{array}{cccc}
e^{-ic_3} c_{-} & 0 & 0 & - i e^{-ic_3} s_{-} \\
0 & e^{ic_3} c_{+} & - i e^{ic_3} s_{+} & 0 \\
0 & -i e^{ic_3} s_{+} & e^{ic_3} c_{+} & 0 \\
-i e^{-ic_3} s_{-} & 0 & 0 & e^{-ic_3} c_{-}
\end{array}
\right),
\end{eqnarray}
\end{widetext}
where $\beta_k$ $(k = 1, 2, 3)$ are real parameters and $\hat{\sigma}_k$ are the Pauli matrices. Here, $c_{\pm} = \cos(\beta_1 \pm \beta_2)$ and $s_{\pm} = \sin(\beta_1 \pm \beta_2)$. Note that the parameters $\beta_k$, often referred to as canonical (or Euler) parameters, can completely characterize the entangling properties of $\hat{U}$. We thus analyze the entangling behavior of all physically-implementable two-qubit gates by EP and EPD of $\hat{d}_\chi$.

 \begin{figure}[t]
 \centering
 \includegraphics[width=0.46\textwidth]{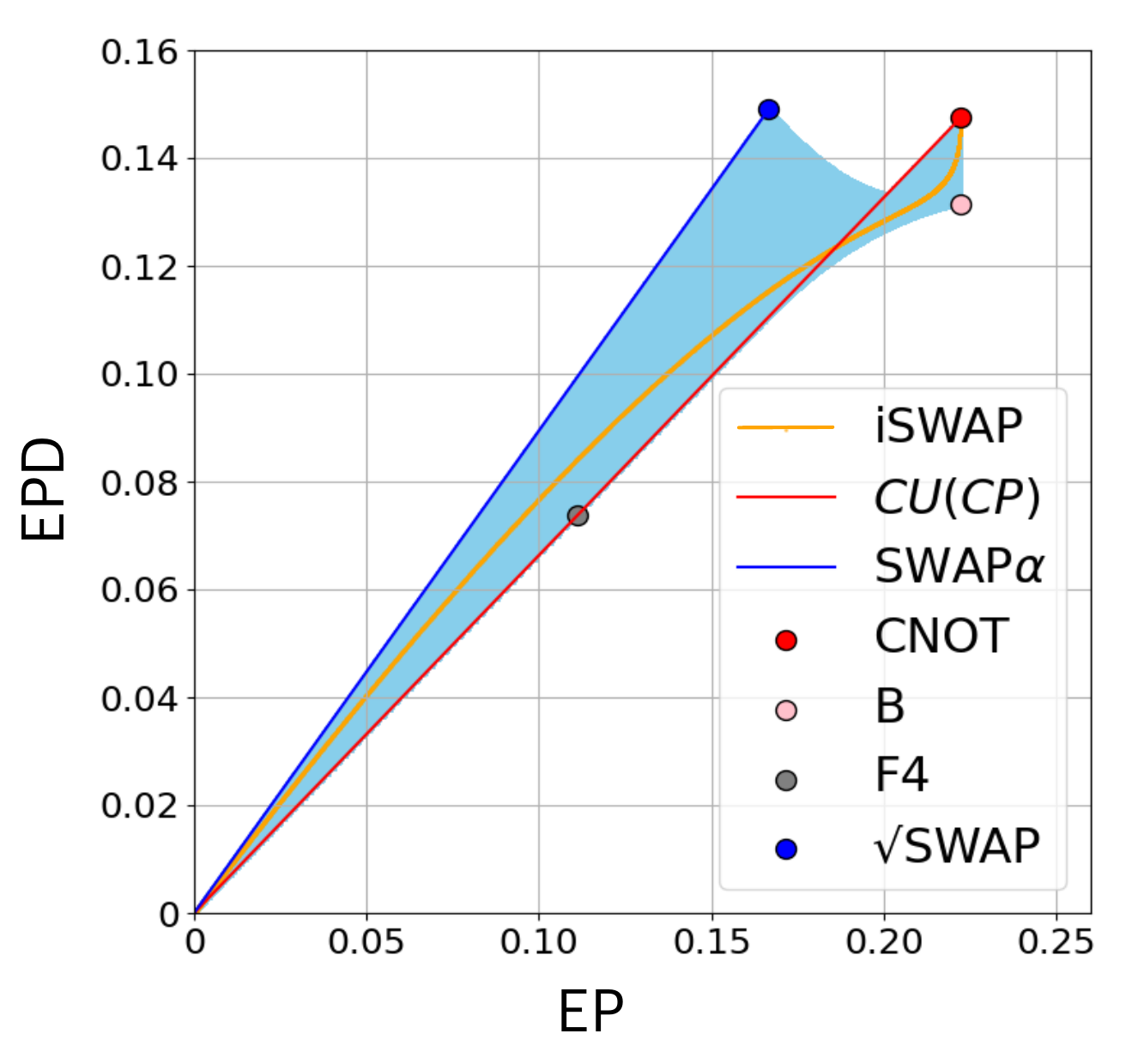}
 \caption{EP–EPD Landscape of Two-Qubit Gates. The shaded region shows the full range of EP and EPD achievable by general two-qubit unitaries. Specific gate families---CU (red), $\mathrm{SWAP}^\alpha$ (blue), and $\mathrm{iSWAP}$ (yellow)---follow characteristic EP--EPD trajectories due to their fixed internal structure. Some representative gates such as CNOT, $B$, $F_4$, and $\sqrt{\mathrm{SWAP}}$ are also indicated. The $B$-gate lies on the maximal EP boundary with moderate EPD, indicating strong yet consistent entanglement generation. The $F_4$-gate resides centrally, achieving moderate EP with low fluctuation, exemplifying a well-balanced entangler. CNOT shows maximal EP but moderate EPD, reflecting selective entangling behavior. In contrast, $\sqrt{\mathrm{SWAP}}$ displays high EPD despite a lower EP, indicating its entangling strength varies widely across inputs.}
 \label{fig:EP-EPD_su4}
\end{figure}

\begin{table}
\centering
\setlength{\tabcolsep}{0.20in}
\renewcommand{\arraystretch}{1.7}
\begin{tabular}{c  c  c  c}
\hline\hline
Gates				& $\left( \beta_1, \beta_2, \beta_3 \right)$				& EP 			& EPD					\\
\hline
 CNOT				& $\left( \frac{\pi}{4}, 0, 0 \right)$					& $\frac{2}{9}$		& $\frac{2\sqrt{11}}{45}$		\\ 
$B$				& $\left( \frac{\pi}{4}, \frac{\pi}{8}, 0 \right)$				& $\frac{2}{9}$		& $\frac{1}{9}\sqrt{\frac{7}{5}}$	\\ 
%SWAP				& $\left( \frac{\pi}{4}, \frac{\pi}{4}, \frac{\pi}{4} \right)$		& $0$			& $0$					\\
 $\sqrt{\textrm{SWAP}}$	& $\left( \frac{\pi}{8}, \frac{\pi}{8}, \frac{\pi}{8} \right)$		& $\frac{1}{6}$		& $\frac{1}{3\sqrt{5}}$		\\
 $F_4$	& $\left( \frac{\pi}{4}, \frac{\pi}{4}, \frac{\pi}{8} \right)$		& $\frac{1}{9}$		& $\frac{\sqrt{11}}{45}$		\\
\hline\hline
\end{tabular}
\caption{The EP and EPD values of representative two-qubit gates are listed. While the CNOT and $B$ gates exhibit the same EP, they differ in EPD, reflecting their distinct entanglement variability.}
\label{tab:2}
\end{table}

The EP of $\hat{d}\chi$ is computed using {\bf Proposition~\ref{prop:2}} as
%\begin{widetext}
\begin{eqnarray}
%e_{\overline{p}_0}(\hat{d}_\chi) &=& \frac{1}{18} \left( 3 - \left( \cos(4\beta_1) \cos(4\beta_2) + \cos(4\beta_2) \cos(4\beta_3) + \cos(4\beta_3) \cos(4\beta_1) \right) \right),
e_{\overline{p}_0}(\hat{d}_\chi) \!\!&=&\!\! \frac{1}{18} \left( 3 - \left( \cos(4\beta_1) \cos(4\beta_2) + \cos(4\beta_2) \cos(4\beta_3) \right. \right. \nonumber \\
&& \quad \left. \left. + \cos(4\beta_3) \cos(4\beta_1) \right) \right),
\end{eqnarray}
%\end{widetext}
achieving its maximum $e_{\overline{p}_0}^\text{max}(\hat{d}_\chi) = \frac{2}{9}$ at $(\beta_1, \beta_2, \beta_3) = \left( \frac{(2k_1 + 1)\pi}{4}, \frac{k_2 \pi}{8}, k_3 \pi \right)$, and minimum zero when $\hat{d}_\chi$ is local products, e.g., at $(\beta_1, \beta_2, \beta_3) = \left( \frac{k_1 \pi }{2}, \frac{k_2 \pi}{2},\frac{k_3 \pi}{2} \right)$ for integers $k_i$. The EPD of $\hat{d}\chi$, from {\bf Proposition~\ref{prop:3}}, is given by
\begin{widetext}
\begin{eqnarray}
\Delta_{\overline{p}_0}(\hat{d}_\chi) &=& \frac{1}{45 \sqrt{2}} \Big[ 57 - 4 \cos(8\beta_2) - 23 \cos(4\beta_2) \cos(4\beta_3) \nonumber \\
&& \quad + \cos(4\beta_1) \Big( (\cos(8\beta_2) - 23) \cos(4\beta_3) + (\cos(8\beta_3) - 23) \cos(4\beta_2) \Big) + \cos(8\beta_3) \left(7 \cos(8\beta_2) - 4 \right) \nonumber \\
&& \quad + \cos(8\beta_1) \Big( 7 \cos(8\beta_2) + \cos(4\beta_2) \cos(4\beta_3) + 7 \cos(8\beta_3) - 4 \Big) \Big]^{\frac{1}{2}}.
\end{eqnarray}
\end{widetext}
with maximum $\Delta_{\overline{p}_0}^\text{max}(\hat{d}_\chi) = \frac{1}{3\sqrt{5}}$ at $(\beta_1, \beta_2, \beta_3) = \left( \frac{(2k_1+1)\pi}{8}, \frac{(2k_2+1)\pi}{8}, \frac{(2k_3+1)\pi}{8} \right)$ and minimum zero at the zero-EP point. The EP and EPD values of several archetypal two-qubit gates (CNOT, $B$, $\sqrt{\textrm{SWAP}}$, and $F_4$) are summarized in Tab.~\ref{tab:2}, with EP values consistent with previous results~\cite{PhysRevResearch.2.043126, PhysRevA.70.052313,PhysRevA.87.022111}. Here, the $B$-gate, specified by the parameters $\left( \frac{\pi}{4}, \frac{\pi}{8}, 0 \right)$, is known as an efficiently implementable perfect entangler~\cite{PhysRevLett.93.020502}. The $F_4$-gate, a unitary of order $4$, is defined by the quantum Fourier transform: $(\hat{F}_4)_{mn} = \frac{1}{2} e^{i \frac{2\pi}{4} mn}$. Notably, this $\mathrm{F}_4$-gate achieves exactly half the EP and EPD of CNOT.

Unlike the CU, $\mathrm{SWAP}^{\alpha}$, and iSWAP gate families---which exhibit a sharp EP--EPD ratio---the general SU($4$) landscape can display a far richer structure. To fully capture this behavior, we sample the SU($4$) space densely over the canonical parameter cube $(\beta_1, \beta_2, \beta_3)$ and plot the resulting EP and EPD values in Fig.~\ref{fig:EP-EPD_su4}. The resulting structure reveals a trade-off surface: two-qubit gates that maximize EP often come at the cost of higher variability, and vice versa. However, the points that maximize EP do not coincide with those that maximize EPD, indicating that there exists no universal two-qubit gate in SU($4$) that simultaneously optimizes both the average and the variability of entanglement generation. This distinction has significant implications: two-qubit entangling gates are not equally effective in all contexts. Thus, a two-qubit gate that consistently produces moderate entanglement across many input states (e.g., $\sqrt{\mathrm{SWAP}}$) may be preferable in tasks requiring uniformity, whereas one that yields strong but input-dependent entanglement (e.g., CNOT) may be more suitable for targeted operations. This asymmetry highlights the importance of EPD as a complementary metric to EP, offering critical information about the distributional characteristics of entanglement generation. This visualization not only situates known gate families within a unified framework but also provides a powerful tool for selecting gates tailored to specific quantum algorithms, depending on whether entanglement strength or uniformity is more critical to performance. To connect this EP--EPD landscape to the underlying distributional picture, we plot the histograms $P_{\hat{U}}(e)$ for several gates in {\bf Appendix~\ref{appendix:dist}}, confirming that gates with the same EP can still have different distribution widths and shapes---precisely what EPD quantifies.

%---------------------------------------------------------------------------------------------------------------------------------------------------------------------------------------------------------
\section{Entangling Power and Entangling Power Deviation of General Bipartite-System Operation in Arbitrary Hilbert Spaces}\label{Sec:5}
%---------------------------------------------------------------------------------------------------------------------------------------------------------------------------------------------------------

\subsection{Systematic reduction of group-theoretic traces}  %-------------------------------------------------------------------------------

In this section, we introduce a generalizable framework to efficiently compute EP and EPD of quantum operations acting on arbitrary Hilbert-space bipartite systems. Typically, the direct evaluation of the group-theoretic expressions, as in {\bf Proposition~\ref{prop:1}}, poses significant computational challenges, especially for bipartite systems with arbitrary Hilbert space dimensions. Specifically, although permutation operators provide powerful tools for exploiting the symmetry structure inherent in multi-copy Hilbert spaces, deriving explicit analytical expressions for EP and EPD remains impractical in general settings involving arbitrary-dimensional subsystems. To address this, we provide a systematic method for reducing complex trace expressions---arising from multiple tensor products of unitary operations---into more tractable group-theoretic forms.

Our method simplifies the trace expressions encountered in EP and EPD computations:
\begin{eqnarray}
\tr \bigl( \hat{U}^{\otimes \kappa} \hat{V}_d(\pi) \hat{V}_d(\sigma) \hat{U}^{\dagger \otimes \kappa}  \hat{V}_d(\alpha) \hat{V}_d(\beta) \bigr),
\end{eqnarray}
$\pi, \nu, \alpha, \beta \in S_{\kappa}$. Employing the composition rules of permutation matrices, as detailed in the properties in Eq.~(\ref{eq:propVd}), this expression simplifies elegantly to:
\begin{eqnarray}
\tr \bigl( \hat{U}^{\otimes \kappa} \hat{V}_d(\pi \circ \nu) \hat{U}^{\dagger \otimes \kappa} \hat{V}_{d}( \pi \circ \nu \circ \alpha \circ \beta) \bigr).
\end{eqnarray}
To further simplify this expression, it is essential to understand how traces involving tensor products of operators and permutation operators behave. This naturally leads to the following fundamental proposition, which allows us to express such traces in terms of disjoint cycle structures within the symmetric group.
\begin{proposition}
Let $\pi \in S_\kappa$, and let $\hat{A}_1, \dots, \hat{A}_\kappa$ be operators represented as $d \times d$ matrices over $\mathbb{C}$. Let $\hat{V}_d(\pi)$ denote the permutation operator acting on the tensor product space $(\mathbb{C}^d)^{\otimes k}$. Then,
\begin{eqnarray}
\mathrm{Tr}\bigl( \hat{A}_1 \otimes \cdots \otimes \hat{A}_\kappa \hat{V}_d(\pi) \bigr) \!=\! \prod_{c \in \pi} \! \mathrm{Tr}\left( \prod_{m=0}^{\kappa_c -1}\!\hat{A}_{c^{-m} (l_c)} \!\!\right)\!, 
\end{eqnarray}
where $\pi = \{ c_1, \dots, c_r \}$ is the disjoint cycle decomposition of $\pi$, each cycle $c = (l_1, \dots, l_{\kappa_c})$ has length $\kappa_c = \abs{c}$, and $l_c$ is an arbitrary reference element of cycle $c$. The notation $c^{-m}(l_c)$ denotes the $m$-th inverse image of $l_c$ under the cycle $c$.
\label{prop:4}
\end{proposition}

\begin{proof}---We first consider the case where $\pi \in S_\kappa$ is a full-length cycle, i.e., $\abs{\pi} = \kappa$. From the action of the permutation operator, we have $\hat{V}_d(\pi) \ket{ j_1, \cdots, j_\kappa} = \ket{ j_{\pi^{-1}(1)}, \cdots, j_{\pi^{-1}(\kappa)} }$. Thus, the trace becomes
\begin{eqnarray}
&& \tr \bigl( \hat{A}_1 \otimes \cdots \otimes \hat{A}_\kappa \hat{V}_d(\pi) \bigr) \nonumber \\
&& \quad = \sum_{j_1, \cdots, j_\kappa}\prod_{l=1}^{\kappa} \bra{j_l} \hat{A}_l \ket{j_{\pi^{-1}(l)}}.
%= \sum_{i_1, \cdots, i_n} \langle i_1 \cdots i_k | A_1 \otimes \cdots  \otimes A_k |i_{\pi^{-1}(1)} \cdots i_{\pi^{-1}(k)} \rangle   \nonumber  \\
\end{eqnarray} 
To simplify this expression, fix an arbitrary position $l_c \in \{ l_1, \dots, l_\kappa \}$. We reorganize the product by tracking the flow of indices under successive applications: i.e., $\pi^{-1}, \pi^{-2}, \dots$, until we return to the starting point (which is guaranteed since $\pi$ is a finite-length cycle). As the inner products $\bra{j_m} \hat{A} \ket{j_n} \in \mathbb{C}$ commute, we can write:
\begin{widetext}
\begin{eqnarray}
\sum_{j_1, \cdots, j_k} \prod_{l=1}^{\kappa} \bra{j_l} \hat{A}_l \ket{j_{\pi^{-1}(l)}} = \sum_{j_1, \cdots, j_\kappa} \bra{j_{l_c}} \hat{A}_{l_c} \ket{j_{\pi^{-1}(l_c)}} \bra{j_{\pi^{-1}(l_c)}} \hat{A}_{\pi^{-1}(l_c)} \ket{j_{\pi^{-2} (l_c)}} \cdots  =  \tr \left( \prod_{l=1}^{\kappa} \hat{A}_{\pi^{-m}(l_c)} \right).
\end{eqnarray}
\end{widetext}
This completes the computation for a single full-length cycle. For a general permutation $\pi \in S_\kappa$, we note that it decomposes into disjoint cycles, and the overall trace factorizes accordingly. For each cycle, the same argument applies, yielding the product over traces corresponding to each cycle. Hence, the result follows.
\end{proof}

This {\bf Proposition~\ref{prop:4}} serves as a key tool in simplifying trace computations involving permutation operators. In particular, it provides a clear combinatorial interpretation of the trace structure in terms of disjoint cycles.

%---------------------------------------------------------------------------------------------------------------------------------------------------------------------------------------------------------
\subsection{EP and EPD of generalized CX operations in arbitrary-dimensional Hilbert spaces}
%---------------------------------------------------------------------------------------------------------------------------------------------------------------------------------------------------------

As an application of {\bf Propositions~\ref{prop:2}, \ref{prop:3}}, and {\bf\ref{prop:4}}, we investigate the analytical behavior of EP and EPD for operations in general bipartite Hilbert spaces. 
%To verify the consistency of our formalism, we first consider a trivial yet illustrative example where no entanglement is generated. 

{\em \textrm{SWAP} operation.}---As a start, we examine the general $(d_1 \times d_2)$-dimensional $\mathrm{SWAP}$ operations, defined as $\mathrm{SWAP}(\ket{\psi_1}\otimes\ket{\psi_2}) = \ket{\psi_2}\otimes\ket{\psi_1}$, with $\ket{\psi_1}$ and $\ket{\psi_2}$ belonging to $d_1$- and $d_2$-dimensional Hilbert spaces, respectively. This example serves as a baseline, clearly demonstrating that our framework naturally captures the absence of entanglement in such symmetric operations. The direct computation shows explicitly that the EP and EPD vanish for the SWAP gate. To verify this, note that for $\hat{U}=\mathrm{SWAP}$, i.e., $\hat{U}= \hat{P}_{12}$, we have the tensor-power relations: $\hat{U}^{\otimes 2} = \hat{P}_{12}\hat{P}_{34}$ and $\hat{U}^{\otimes 4} = \hat{P}_{12}\hat{P}_{34}\hat{P}_{56}\hat{P}_{78}$. Applying the composition property of permutation operators (as in Eq.~(\ref{eq:propVd})), specifically, $\hat{V}_{d}(\pi)\hat{V}_{d}(\nu)=\hat{V}_{d}(\pi\circ\nu)$, we obtain the commutation relations:
\begin{eqnarray}
&& \bigl[ \hat{P}_{12} \hat{P}_{34},  \hat{P}^{+}_{13} \hat{P}^{+}_{24} \bigr] =0 \nonumber \\
&& \bigl[ \hat{P}_{12} \hat{P}_{34} \hat{P}_{56} \hat{P}_{78},  \hat{P}^{+}_{1357} \hat{P}^{+}_{2468} \bigr] =0.
\end{eqnarray}
These commutations directly imply that both EP and EPD vanish for the SWAP operation. While trivial, this example provides an essential consistency check and explicitly clarifies the conditions under which EP and EPD vanish for general bipartite Hilbert spaces.

{\em Generalized CX operation.}---We next analyze a natural generalization of the CU family, which we refer to as the generalized CX operation. This family is defined by controlled applications of mutually orthogonal unitaries, and is inspired by the structure involving powers of unitary operators~\cite{PhysRevA.62.030301}. This formulation enables closed-form expressions for both EP and EPD. Interestingly, the behavior of EP and EPD under this class reveals rich dimension-dependent structures, including pronounced parity effects between even and odd dimensions, and allows us to analytically explore the asymptotic scaling of the EP-to-EPD ratio in entanglement generation. For simplicity, we restrict our analysis to the symmetric case where $d_1 = d_2 = d$.
\begin{definition}
The generalized CX operation is
\begin{eqnarray}
\hat{U}_\text{CX} = \sum_{\alpha=0}^{d-1} \hat{\Pi}_{\alpha} \otimes \hat{V}_{\alpha},
\label{eq:CX_def}
\end{eqnarray}
where each $\hat{V}_\alpha$ ($\alpha = 0, 1, \ldots, d-1$) is a local unitary operator with the powers of some $\hat{X}$ matrices, i.e., $\hat{V}_{\alpha} = \hat{X}^{\alpha}$, where $\hat{X}$ is the qudit shift operator defined by $\hat{X}\ket{j}=\ket{j+1\ (\mathrm{mod}\ d)}$ ($j=0,\ldots,d-1$). Here, the following orthogonality condition is satisfied:
\begin{eqnarray}
\mathrm{Tr}\bigl( \hat{V}_{\alpha}^{\dagger} \hat{V}_{\beta} \bigr) = d \delta_{\alpha \beta},
\label{eq:CX_condi}
\end{eqnarray}
where $\hat{\Pi}_\alpha$ is an orthonormal rank-one projector satisfying: $\mathrm{Tr}\bigl( \hat{\Pi}_{\alpha} \bigr) = 1$, $\hat{\Pi}_{\alpha} \hat{\Pi}_{\beta} = \delta_{\alpha \beta} \hat{\Pi}_{\beta}$, $\hat{\Pi}_{\alpha} = \hat{\Pi}_{\alpha}^{\dagger}$, and $\sum_{\alpha=0}^{d-1} \hat{\Pi}_{\alpha} = \hat{\openone}$.
\end{definition}
This class of quantum operations acts on the bipartite Hilbert space ${\cal H}^d \otimes {\cal H}^d$ and is capable of generating genuine bipartite entanglement. For example, applying $\hat{U}_\text{CX}$ to a product state of the form $\sum_{j=0}^{d-1} \ket{j} \otimes \ket{0}$ can yield a maximally entangled state such as $\sum_{j=0}^{d-1} \ket{j} \otimes \ket{j}$. In the case of $d=2$, this generalized CX gate reduces exactly to the conventional CU gate family, discussed in Sec.~\ref{Sec:4}.

We now evaluate EP of the generalized CX operation. To this end, we compute the four constituent terms of EP by utilizing the structural conditions in Eq.~(\ref{eq:CX_condi}) and the algebraic properties of $\hat{\Pi}_\alpha$. Specifically, we find:
\begin{eqnarray}
&\mathrm{(i)}& \tr \bigl( \hat{P}_{13}^{-} \bigr)  = \frac{1}{2} \left( d^4 -d^3 \right) \nonumber \\
&\mathrm{(ii)}& \tr \bigl( \hat{U}^{\otimes 2} \hat{P}_{13} \hat{P}_{24} \hat{U}^{\dagger \otimes 2} \hat{P}_{13}^{-} \bigr) = - \frac{1}{2} \left(d^3 - d^2\right) \nonumber \\
&\mathrm{(iii)}& \tr \bigl( \hat{U}^{\otimes 2} \hat{P}_{13} \hat{U}^{\dagger \otimes 2} \hat{P}_{13}^{-} \bigr) = \frac{1}{2} \bigl( d^3 \!-\! d^2 \sum_{\alpha} \abs{\tr\hat{\Pi}_{\alpha}}^2 \bigr) \!=\! 0\nonumber \\
&\mathrm{(iv)}& \tr \bigl( \hat{U}^{\otimes 2} \hat{P}_{24} \hat{U}^{\dagger \otimes 2} \hat{P}_{13}^{-} \bigr) = \frac{1}{2} (d^3 - d^2).
\end{eqnarray}
Summing all contributions, we obtain the compact expression for the EP:
\begin{eqnarray}
e_{\overline{p}_0}(\hat{U}_\text{CX}) = \frac{d^2}{(d+1)^2} E(\hat{U}_\text{CX}) = \frac{d(d-1)}{(d+1)^2},
\end{eqnarray}
where 
\begin{eqnarray}
E(\hat{U}_\text{CX}) = 1 - \frac{1}{d^2} \sum_{\alpha} \abs{\tr(\Pi_{\alpha})}^2 = 1 - \frac{1}{d}.
\end{eqnarray}
This result matches the earlier findings in Ref.~\cite{PhysRevA.62.030301}.

The computation of EPD, however, is more involved due to the presence of higher-order moments. According to {\bf Proposition~\ref{prop:3}}, the EPD computation involves the evaluation of several intricate trace terms, such as
\begin{widetext}
\begin{eqnarray}
\sum_{\alpha_1, \alpha_2, \alpha_3, \alpha_4}  \tr \bigl( \hat{V}_{\alpha_1} \hat{V}_{\alpha_2}^{\dagger} \hat{V}_{\alpha_3} \hat{V}_{\alpha_1}^{\dagger} \hat{V}_{\alpha_2} \hat{V}_{\alpha_4}^{\dagger} \hat{V}_{\alpha_3} \hat{V}_{\alpha_4}^{\dagger} \bigr) \quad\text{and}\quad \sum_{\alpha_1, \alpha_2, \alpha_3, \alpha_4} \abs{\tr \bigl( \hat{V}_{\alpha_1} \hat{V}_{\alpha_2}^{\dagger} \hat{V}_{\alpha_3} \hat{V}_{\alpha_4}^{\dagger} \bigr)}^2.
\end{eqnarray}
\end{widetext}
Fortunately, using the specific structure of $\hat{U}_\text{CX}$ and the simplifying identities in Eq.~(\ref{eq:CX_condi}), these trace expressions can be reduced analytically. Applying {\bf Proposition\ref{prop:4}}, we derive the following closed forms:
\begin{widetext}
\begin{eqnarray}
F_{id} &=& d^2(d+1)^2 (d+2)^2(d+3)^2 \nonumber \\[2pt]
F_{(13)} &=& d^2  (d+2)^2(d+3)^2 (1+3d) \nonumber \\
F_{(13)(57)} &=&
\begin{cases}
d^2 \left( 26 + d \left( 200+d(247+d(86+9d) \right) \right), \quad d=\textrm{even} \\[2pt]
d^2 \left( 36 + d \left( 198+d(247+d(86+9d) \right) \right), \quad d=\textrm{odd}
\end{cases}
\end{eqnarray}
\end{widetext}
The difference in $F_{(13)(57)}$ between even and odd $d$ arises from the parity-dependent trace value of
\begin{eqnarray}
 \sum_{\alpha_1, \alpha_2} \abs{ \tr \bigl( \hat{V}_{\alpha_1} \hat{V}_{\alpha_2}^{\dagger} \hat{V}_{\alpha_1} \hat{V}_{\alpha_2}^{\dagger} \bigr) }^2
 \label{eq:trVVVV},
\end{eqnarray}
which depends on the number of integer solutions to the congruence $2(\alpha_1 - \alpha_2) \equiv 0~(\text{mod}~d)$. This yields $d$ solutions for odd $d$ and $2d$ for even $d$, resulting in trace sums of $d^3$ and $2d^3$, respectively. Since this term appears twice within the $(4!)^2$ terms contributing to $F_{(13)(57)}$, the overall discrepancy between even and odd dimensions is $2d^3$. With these in hand, we have
\begin{widetext}
\begin{eqnarray}
&& 4 (4!)^2 D_{d}^{2} \tr \bigl( \hat{U}^{\otimes 4} \hat{P}_{1357}^{+} \hat{P}_{2468}^{+} \hat{U}^{\dagger \otimes 4} \hat{P}_{13}^{-} \hat{P}_{57}^{-} \bigr)  =
\begin{cases}
\frac{ d^6+6 d^5+5 d^4-12 d^3+6 d^2-4 d}{(d+1)^2(d+2)^2(d+3)^2}, \quad d=\textrm{even} \\
\\
\frac{ d^6+6 d^5+5 d^4-12 d^3+6 d^2-6 d}{(d+1)^2(d+2)^2(d+3)^2}, \quad d=\textrm{odd}
\end{cases} \nonumber 
\end{eqnarray}
\end{widetext}
Substituting these results into the full expression for EPD, we obtain the closed-form:
\begin{eqnarray}
\Delta_{\overline{p}_0}(\hat{U}_\text{CX}) =
\left\{
\begin{array}{ll}
\Theta_{\textrm{even}} , & \textrm{$d$ even} \\
\Theta_{\textrm{even}} - {\cal K}, & \textrm{$d$ odd}
\end{array} 
\right.
\end{eqnarray}
where 
\begin{eqnarray}
\Theta_{\textrm{even}} &=& \frac{8 d^5+34 d^4+8 d^3-20 d^2-4 d}{(d+1)^4(d+2)^2(d+3)^2} \nonumber \\
\mathcal{K} &=& \frac{2d( d^2+ 11d +1)}{(d+1)^4(d+2)^2(d+3)^2}.
\end{eqnarray}
Through this analysis, we demonstrate that the entanglement generation patterns arising from the dimension-dependent parity in generalized CX operations cannot be distinctly captured by EP alone. In contrast, EPD effectively reveals subtle differences in the entanglement generation associated with the dimensional parity, highlighting its role as a more refined diagnostic measure. Thus, it is confirmed that an accurate characterization of entangling behavior requires the combined analysis of both EP and EPD.

%---------------------------------------------------------------------------------------------------------------------------------------------------------------------------------------------------------
\section{Summary}
%---------------------------------------------------------------------------------------------------------------------------------------------------------------------------------------------------------

In this work, we have developed a unified and rigorous framework for analyzing the entangling behavior of quantum operations based on two complementary measures: the entangling power (EP) and the entangling power deviation (EPD). While EP quantifies the average amount of entanglement that a unitary operation can generate from all possible bipartite product input states, EPD captures the fluctuation of this entanglement across different inputs, thereby reflecting the input-state sensitivity or bias of the entanglement generation. Together, these two quantities provided a more complete and physically meaningful characterization of quantum entangling behavior---not only in strength but also in consistency. We began by deriving closed-form expressions for both EP and EPD using a group-theoretical formalism grounded in Schur-Weyl duality. By expressing Haar-averaged moments in terms of the permutation operators, we revealed how the entangling characteristics of unitaries are governed by their symmetry properties and algebraic structure. Notably, we presented general expressions that are applicable to arbitrary bipartite Hilbert-space dimensions, providing a robust foundation for analyzing the entangling behaviors. In particular, our moment-operator formulation recovers the EP analysis of Ref.~\cite{PhysRevA.62.030301} as the $\kappa=2$ case and provides a systematic route to higher moments---most notably the $\kappa=4$ moment required for EPD and, in principle, arbitrary $\kappa$th for higher-order statistics of entanglement generation.

In the analyses, we identified the precise algebraic conditions under which a unitary operation fails to generate entanglement. In {\bf Theorem~\ref{thm:1}}, we showed that EP vanishes when the unitary operation commutes with certain symmetric projectors in its two-copy tensor representation. This condition encompasses both product unitaries---such as $\hat{U} = \hat{u}_1 \otimes \hat{u}_2$---and non-product but non-entangling gates, such as the SWAP operation. We then established in {\bf Theorem~\ref{thm:2}} that EPD vanishes if and only if EP vanishes. Furthermore, in {\bf Corollary~\ref{corol:1}} we provided an operational interpretation of EPD as the susceptibility of the mean entanglement under an exponential biasing of the input ensemble, clarifying how EPD quantifies the input-state sensitivity beyond EP. This confirms that entanglement generation and its input-state sensitivity are intrinsically linked. Such an intrinsic link has, to our knowledge, not been explicitly characterized before, and opens a new perspective on the nature of quantum entanglement generation.

To demonstrate the utility of this framework, we applied it to several representative families of two-qubit gates, including controlled-unitary (CU), $\mathrm{SWAP}^\alpha$, and $\mathrm{iSWAP}$ gates, as well as general SU($4$) unitaries. For the CU and $\mathrm{SWAP}^\alpha$ families, both EP and EPD admit closed-form expressions and vary proportionally with gate parameters, yielding constant EP-to-EPD ratios. Notably, the ratio is larger for $\mathrm{SWAP}^\alpha$ gates, indicating stronger input-state sensitivity despite lower average entanglement. In contrast, the $\mathrm{iSWAP}$ family exhibits a nonlinear EP--EPD relationship. Although its maximal EP and EPD match those of the CU gates, the non-uniform entanglement distribution reflects a richer dependence on input states. This difference underscores that the two-qubit gates with similar EP can exhibit fundamentally distinct entangling behaviors---something captured only through EPD. Extending the analysis to the full SU($4$) space, we visualized the global EP--EPD structure using the KAK decomposition. This affirmed the nontrivial trade-off landscape: the gates maximizing EP do not coincide with those maximizing EPD, and no two-qubit gate optimizes both. These findings confirm our central result: entanglement generation inevitably exhibits input-state dependence, and EP must be complemented by EPD to characterize both strength and variability (see {\bf Theorem~\ref{thm:1}} and {\bf \ref{thm:2}}). This analysis positions EPD not as a secondary refinement, but as a fundamental metric that complements EP.

Extending the analysis to higher-dimension, we considered a class of generalized controlled-unitary---called generalized CX operations---acting on ${\cal H}^{d} \otimes {\cal H}^{d}$. We derived exact closed-form expressions for their EP and EPD, revealing a rich dimension-dependent structure in the entanglement generation. Notably, we found that these gates exhibit parity-dependent behavior---that is, the entanglement fluctuation differs qualitatively depending on whether the system dimension $d$ is even or odd. This subtle effect is entirely invisible to EP but is captured faithfully by EPD, highlighting its diagnostic power. These results underscore (again) the necessity of EPD as a second-order metric that complements EP and reveals deeper structural features of entanglement generation.

Overall, our findings not only offer analytic benchmarks but also provide conceptual insights into the fundamental behaviors of entanglement generation. By positioning EPD as an essential component in the analysis of entangling behavior, this work lays the groundwork for future studies in both theory and application. Possible directions include extending EP and EPD to multipartite systems~\cite{2410.03361,PhysRevA.67.042323} and ancilla-assisted circuits~\cite{PhysRevA.67.042323}, exploring analogous measures based on different entanglement monotones, and analyzing the robustness of entangling behavior under noise, decoherence, or open-system dynamics. In addition, understanding how EP and EPD relate to performance in practical quantum protocols---such as variational quantum algorithms and/or quantum neural networks~\cite{Sim2019expressibility,Cerezo2022challenges,Kim2022quantum}---may provide principled criteria for optimizing the tasks under practical constraints.

In conclusion, EP and its deviation, EPD, together form a comprehensive and physically grounded framework for understanding how quantum operations generate entanglement. The introduction and analysis of EPD reveal that the input-state dependence of entanglement generation is in fact a fundamental and unavoidable aspect of quantum information processing.

%---------------------------------------------------------------------------------------------------------------------------------------------------------------------------------------------------------
\section*{Acknowledgement}
%---------------------------------------------------------------------------------------------------------------------------------------------------------------------------------------------------------

JB thanks Dr.~Antonio Mandarino and Prof.~Marek {\.Z}ukowski for discussions. This work was supported by the Ministry of Science, ICT and Future Planning (MSIP) by the National Research Foundation of Korea (RS-2024-00432214, RS-2025-03532992, RS-2023-NR119931, and RS-2025-18362970) and the Institute of Information and Communications Technology Planning and Evaluation grant funded by the Korean government (RS-2019-II190003, ``Research and Development of Core Technologies for Programming, Running, Implementing and Validating of Fault-Tolerant Quantum Computing System''), the Korean ARPA-H Project through the Korea Health Industry Development Institute (KHIDI), funded by the Ministry of Health \& Welfare, Republic of Korea (RS-2025-25456722), and the Ministry of Trade, Industry, and Energy (MOTIE), Korea, under the project ``Industrial Technology Infrastructure Program'' (RS-2024-00466693). This work is also supported by the Grant No.~K25L5M2C2 at the Korea Institute of Science and Technology Information (KISTI).

\onecolumngrid

%-------------------------------
\appendix
%-------------------------------

%---------------------------------------------------------------------------------------------------------------------------------------------------
\section{Theoretical Foundations of Haar-Averages and Moment Operators}\label{appendix:A}
%---------------------------------------------------------------------------------------------------------------------------------------------------

This appendix summarizes several essential definitions and fundamental theorems underpinning the derivation of EP and EPD. The derivation presented in {\bf Proposition~\ref{prop:1}}, {\bf Proposition~\ref{prop:2}} and {\bf Proposition~\ref{prop:3}} relies on standard techniques from the Haar measure on the unitary group and its associated moment operators. These results follow from the Schur-Weyl duality and the modern framework of Weingarten calculus~\cite{Collins:2003ncs,Collins:2006jgn}. We begin by recalling the definition of the Haar measure and introducing the notion of $\kappa$-th moment operators, which describe the averaged action of random unitaries on $\kappa$-fold tensor product spaces. We then explain the algebraic structure of these operators using the commutant algebras characterized by the Schur-Weyl duality. Finally, we state key formulas for the Haar averages of tensor powers of pure states, which play a central role in expressing quantities such as $\hat{\Omega}_{\overline{p}_0}^{(2)}$ and $\hat{\Omega}_{\overline{p}_0}^{(4)}$ appearing in our EP and EPD analysis.

Rather than providing full technical proofs, this appendix emphasizes the essential statements of the relevant theorems and propositions, together with remarks highlighting their significance and applications in the context of Haar-random product states and its potential applications in quantum information theory. For a more technical derivation and detailed proofs of the results presented below, see the comprehensive reviews in Refs.~\cite{Zhang:2014zbq, Ragone:2022axl, Mele2024introductiontohaar}.

%---------------------------------------------------------------------------------------------------------------------------------------------------------------------
\subsection{Haar measure}

\begin{definition}
The Haar measure $\mu_H$ for unitary $U(d)$ is a probability measure satisfying  
\begin{eqnarray}
&& \int_S d\mu_{H}(\hat{U}) \geq 0, \quad \forall S \subset U(d), \nonumber \\
&& \int_{U(d)} d\mu_H(\hat{U}) = 1.
\end{eqnarray}
For any function $f(\hat{U})$ over the Haar measure, one defines its expectation value, or the Haar integral as
\begin{eqnarray}
\mathbb{E}_{\hat{U} \sim \mu_H} [f(\hat{U})] := \int_{U(d)} f(\hat{U}) \, d \mu_H(\hat{U}).
\end{eqnarray}
\end{definition}

\begin{remark}
The Haar measure defines the natural uniform distribution over the unitary group and underpins average calculations involving random unitaries in quantum information theory. It plays a central role, e.g., in modeling quantum unitary design~\cite{Gross2007evenly,Ambainis2007quantum,Roberts2017chaos}, analyzing entanglement~\cite{Nahum2017quantum,Gomes2023entanglement}, averaging the fidelities~\cite{Bang2018fidelity}, quantifying the expressibility of random circuits~\cite{Sim2019expressibility}, etc.
\end{remark}

\begin{proposition}
The Haar measure provides the unique uniform probability distribution over the unitary group, satisfying 
\begin{itemize}
\item left and right invariance: for all $\hat{U}, \hat{V} \in U(d)$, 
\begin{eqnarray}
\int f(\hat{V}\hat{U}) \, d\mu_H(\hat{U}) &=& \int f(\hat{U}) \, d\mu_H(\hat{U}), \\
\int f(\hat{U}\hat{V}) \, d\mu_H(\hat{U}) &=& \int f(\hat{U}) \, d\mu_H(\hat{U}).
\end{eqnarray}
\item  invariant under conjugation and self-adjointness under inversion 
\begin{eqnarray}
\int f(\hat{V}\hat{U}\hat{V}^{\dagger}) d \mu_H (\hat{U}) &=& \int f(\hat{U}) d\mu_H (\hat{U}), \\
\int f(\hat{U}^\dagger) d\mu_H(\hat{U}) &=& \int f(\hat{U}) d\mu_H(\hat{U}).
\end{eqnarray}
\end{itemize}
\end{proposition}
These properties make the Haar integral a natural choice for averaging over unitaries.

\begin{remark}
The Haar measure is uniquely characterized by its invariance properties and exists on all compact groups, including $U(d)$. It serves as a foundational tool in the representation theory of compact groups and in random matrix theory.
\end{remark}

%---------------------------------------------------------------------------------------------------------------------------------------------------------------------
\subsection{$\kappa$-th moment operators}

Next, to study how random unitaries act on multi-copy systems, we introduce the notion of $\kappa$-th moment operators, which describe the average action of $\kappa$-fold tensor powers of unitary matrices over the Haar measure.
\begin{definition}[$\kappa$-th Moment operator] 
Let $\kappa \in \mathbb{N}$. The $\kappa$-th moment operator $\mathcal{M}_{\mu_H}^{(\kappa)} : \mathcal{L}((\mathbb{C}^d)^{\otimes \kappa}) \rightarrow \mathcal{L}((\mathbb{C}^d)^{\otimes \kappa})$ is defined as
\begin{eqnarray}
\mathcal{M}_{\mu_H}^{(\kappa)}(\mathcal{\hat{O}}) := \mathbb{E}_{\hat{U} \sim \mu_H} \bigl[ \hat{U}^{\otimes \kappa} \mathcal{\hat{O}} \, \hat{U}^{\dagger \otimes \kappa} \bigr]
\end{eqnarray}
for all $\mathcal{\hat{O}} \in \mathcal{L} ((\mathbb{C}^d)^{\otimes \kappa})$. Then, we provide the following remarks:
\end{definition}

\begin{remark}
The $\kappa$-th moment operator captures the average action of conjugating an operator $\mathcal{\hat{O}}$ by $\kappa$-fold tensor powers of Haar-random unitaries. Its structure reflects the permutation symmetry of multi-copy Hilbert spaces and is fully characterized by the commutant algebra of $U(d)^{\otimes \kappa}$.
\end{remark}

\begin{remark}
The $\kappa$-th moment operator is linear, trace-preserving, and self-adjoint with respect to the Hilbert-Schmidt inner product. It acts as an orthogonal projector onto the commutant algebra of $U(d)^{\otimes \kappa}$---the subspace of operators commuting with all $\kappa$-fold tensor powers of unitaries---thereby, significantly simplifying the computation of Haar averages within this symmetric subspace.
\end{remark}

\begin{remark}
The $\kappa$-th moment operator also plays a central role in the theory of unitary $t$-designs. A finite ensemble of unitaries $\{\hat{U}_j\}$ forms an exact unitary $t$-design if the average of $\kappa$-fold tensor conjugations over the ensemble matches the Haar moment operator for all $\kappa \leq t$, i.e., if
\begin{eqnarray}
\frac{1}{|\mathcal{E}|} \sum_{\hat{U}_j \in \mathcal{E}} \hat{U}_j^{\otimes \kappa} \mathcal{\hat{O}} \, \hat{U}_j^{\dagger \otimes \kappa} 
= \mathcal{M}_{\mu_H}^{(\kappa)}(\mathcal{\hat{O}}).
\end{eqnarray}
This equivalence means that the statistical properties of Haar-random unitaries up to $\kappa$-th moments can be faithfully reproduced by a finite set of unitaries~\cite{Seymour1984averaging,Dankert2009exact,Ketterer2020entanglement}.
\end{remark}

%---------------------------------------------------------------------------------------------------------------------------------------------------------------------
\subsection{Schur-Weyl duality and Haar-average}

To characterize the structure of the $\kappa$-th moment operator, it is crucial to understand the set of operators that commute with all $\kappa$-fold tensor powers of unitary operators. This set forms the commutant algebra, whose structure is elegantly captured by Schur-Weyl duality. We now state this fundamental result, which provides an explicit description of the commutant associated with the $\kappa$-fold tensor representation of the unitary group.

Firstly, we provide the definition of $\kappa$-th order commutant:
\begin{definition}
Given $S \subset \mathcal{L}(\mathbb{C}^d)$, one defines the $\kappa$-th order commutant as
\begin{eqnarray}
\textrm{Comm}(S,\kappa) := \left\{ \hat{A} \in \mathcal{L}((\mathbb{C}^d)^{\otimes \kappa}) ~:~ \bigl[ \hat{A}, \hat{B}^{\otimes \kappa} \bigr] = 0, \; \forall \hat{B} \in S \right\},
\end{eqnarray}
which forms a vector subspace of $\mathcal{L}((\mathbb{C}^d)^{\otimes \kappa})$.
\end{definition}

\begin{remark}
When $S = U(d)$, the commutant algebra $\mathrm{Comm}(U(d),\kappa)$ admits a remarkably simple structure due to Schur-Weyl duality. This duality captures the full symmetry of $\kappa$-fold tensor product spaces and shows that the commutant is generated by the symmetric group $S_\kappa$, acting through permutation operators.
\end{remark}

We then state the theorem of Schur-Weyl duality as follows.
\begin{theorem}[Schur-Weyl duality]
The $\kappa$-th order commutant of the unitary group is the span of the permutation operators associated with the symmetric group $S_\kappa$:
\begin{eqnarray}
\textrm{Comm}(U(d),\kappa) = \textrm{span}\left( \hat{V}_d(\pi) : \pi \in S_\kappa \right).
\end{eqnarray}
\label{thm:3A}
\end{theorem}

\begin{remark}
Schur-Weyl duality shows that the only operators invariant under conjugation by $\kappa$-fold tensor powers of unitaries are linear combinations of permutation operators. This fundamental fact underlies the Weingarten calculus, which expresses Haar averages as the sums over permutations weighted by the Weingarten function~\cite{Collins:2003ncs,Collins:2006jgn}.
\end{remark}

Using the Schur-Weyl duality and the properties of Weingarten function, the $\kappa$-th moment operator can be expressed as a double sum over permutations. This is stated in the theorem:
\begin{theorem}
The $\kappa$-th moment operator has the following expansion:
\begin{eqnarray}
\mathcal{M}_{\mu_H}^{(\kappa)} (\mathcal{\hat{O}}) = \sum_{\pi, \nu \in S_\kappa} \mathrm{Wg}(\pi^{-1}\nu, d) \; \mathrm{tr}\bigl( \hat{V}_d^\dagger(\nu) \mathcal{\hat{O}} \bigr) \hat{V}_d(\pi),
\end{eqnarray}
where $\mathrm{Wg}(\cdot, d)$ is the Weingarten function associated with $U(d)$.
\label{thm:4A}
\end{theorem}

\begin{remark}
This formula enables explicit computation of Haar averages for multi-copy operators by expressing the average as a weighted sum over permutations. It provides a practical and powerful tool for evaluating quantities such as $\hat{\Omega}_{\overline{p}_0}^{(2)}$ and $\hat{\Omega}_{\overline{p}_0}^{(4)}$ in {\bf Proposition~\ref{prop:2}}, which are central to analyzing EP and EPD.
\end{remark}

As a direct application of {\bf Theorem~\ref{thm:4A}}, we consider the Haar average of tensor powers of a pure state---an essential ingredient in deriving the explicit forms of the second and fourth moment operators for random product states:
\begin{proposition}
For any $|\phi\rangle \in \mathbb{C}^d$, 
\begin{eqnarray}
\mathbb{E}_{\hat{U} \sim \mu_H} \bigl[ \hat{U}^{\otimes \kappa} \ket{\phi}\bra{\phi}^{\otimes \kappa} \hat{U}^{\dagger \otimes \kappa} \bigr] = \mathbb{E}_{|\psi\rangle \sim \mu_H} \bigl[ \ket{\phi}\bra{\phi}^{\otimes \kappa} \bigr] = \frac{\hat{P}_{\mathrm{sym}}^{(d,\kappa)}}{\mathrm{tr}\hat{P}_{\mathrm{sym}}^{(d,\kappa)}},
\end{eqnarray}
where $\hat{P}_{\mathrm{sym}}^{(d,\kappa)} = \frac{1}{\kappa!} \sum_{\pi \in S_\kappa} \hat{V}_d(\pi)$ is the projector onto the symmetric subspace of $(\mathbb{C}^d)^{\otimes \kappa}$ and its trace is given by $\mathrm{tr}\hat{P}_{\mathrm{sym}}^{(d,\kappa)} = \binom{\kappa+d-1}{\kappa}$.
\end{proposition}

\begin{remark}
This result follows directly from the Weingarten expansion, utilizing the permutation invariance of tensor power states---namely, $\ket{\phi}\bra{\phi}^{\otimes \kappa}$ remains unchanged under permutation operators. It yields the exact form of the averaged tensor power states, which define the moment operators $\hat{\Omega}_{\overline{p}_0}^{(2)}$ and $\hat{\Omega}_{\overline{p}_0}^{(4)}$ in our analysis. These expressions form the basis for evaluating entangling power and its deviation over symmetric subspaces.
\end{remark}

%-------------------------------------------------------------------------------------------------------------------------------------------------------------------------------------------------------------------------------------
\section{Entanglement-generation distribution induced by Haar-random product inputs}\label{appendix:dist}
%-------------------------------------------------------------------------------------------------------------------------------------------------------------------------------------------------------------------------------------

Here we provide a qualitative information about the full distribution of entanglement generation induced by a fixed unitary $\hat{U}$ when the input is a Haar-random product state. This complements the moment-based indicators in the main text: EP and EPD summarize the mean and standard deviation of this distribution, but they do not uniquely determine its entire shape.

{\em Definition and relation to EP $\&$ EPD.}---For a fixed unitary $\hat{U}$, the mapping $(\psi_1,\psi_2) \mapsto e=E(\hat{U}\ket{\psi_1}\otimes\ket{\psi_2})$ induces a probability distribution of the entanglement values:
\begin{eqnarray}
P_{\hat{U}}(e) := \int d\mu(\psi_1,\psi_2) \delta \left(e - E(\hat{U}\ket{\psi_1}\otimes\ket{\psi_2})\right).
\end{eqnarray}
By construction, $e_p(\hat{U}) = \int de\, e\,P_{\hat{U}}(e)$ and $\Delta_p(\hat{U})^2=\int de\, e^2 P_{\hat{U}}(e)-e_p(\hat{U})^2$.
Since the linear entropy $E$ is bounded, $P_{\hat{U}}(e)$ cannot be heavy-tailed in the strict probabilistic sense, but it can still be strongly skewed or sharply peaked, and in principle may exhibit features, such as, the multimodality.

\begin{table}[t]
\centering
\setlength{\tabcolsep}{0.22in}
\renewcommand{\arraystretch}{2.0}
\begin{tabular}{c  c  c  c  c}
\hline\hline
Gate & analytic EP & MC mean & analytic EPD & MC std \\
\hline
CNOT & $\frac{2}{9}$ & $0.222$ & $\frac{2\sqrt{11}}{45}$ & $0.148$ \\
$B$ & $\frac{2}{9}$ & $0.222$ & $\frac{1}{9}\sqrt{\frac{7}{5}}$ & $0.132$ \\
$\sqrt{\mathrm{SWAP}}$ & $\frac{1}{6}$ & $0.166$ & $\frac{1}{3\sqrt{5}}$ & $0.149$ \\
\hline\hline
\end{tabular}
\caption{Monte Carlo estimates of the first two moments of $P_{\hat{U}}(e)$ (mean and standard deviation) for $N=2\times 10^5$ Haar-random product inputs per gate, compared with the analytical EP and EPD values (see Tab.~\ref{tab:2} in the main text).}
\label{tab:dist_moments}
\end{table}

\begin{figure}[t]
\centering
\includegraphics[width=1.00\textwidth]{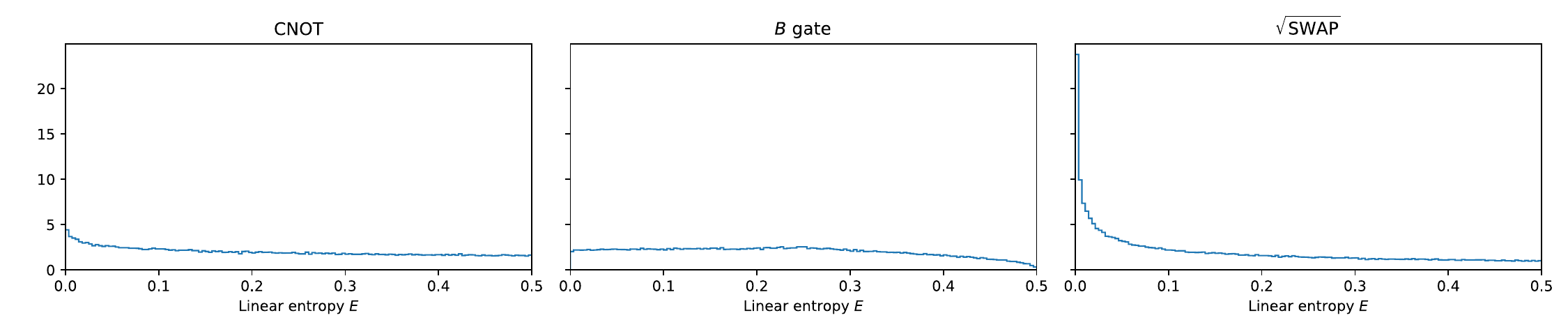}
\caption{The histogram of the generated entanglement $e=E(\hat{U}\ket{\psi_1}\otimes\ket{\psi_2})$ for Haar-random product inputs ($N=2 \times 10^5$ samples per gate), shown for CNOT, $B$ (a perfect entangler with the same EP as CNOT), and $\sqrt{\mathrm{SWAP}}$. The linear entropy $e$ is bounded ($0 \le e \le 1/2$ for two-qubit systems), and the distributions are well-behaved yet exhibit distinct skewness and widths.}
\label{fig:dist_hist_twoqubit}
\end{figure}

{\em Numerical sampling.}---To visualize $P_{\hat{U}}(e)$, we generate the histograms by Monte Carlo (MC) sampling of Haar-random product inputs: (i) we draw $\ket{\psi_1}$ and $\ket{\psi_2}$ independently from the single-qubit Haar measure, (ii) apply $\hat{U}$, (iii) compute $e=E(\hat{U}\ket{\psi_1}\otimes\ket{\psi_2})$, and then (iv) estimate the density of $P_{\hat{U}}(e)$ by binning the samples~\footnote{Because the input ensemble is Haar-invariant and the entanglement measure is invariant under local unitaries, the resulting distribution depends only on the local-equivalence class of $\hat{U}$ (i.e., it is unchanged under $\hat{U} \mapsto (\hat{u}_1\otimes\hat{u}_2)\hat{U}(\hat{v}_1\otimes\hat{v}_2)$).} For the representative two-qubit gates (CNOT, $B$, $\sqrt{\mathrm{SWAP}}$), we use $N = 2 \times 10^5$ random product inputs per gate. As an additional consistency check of the analytical derivations in the main text (which rely on the exact second and fourth Haar moments), the MC mean and standard deviation agree with the closed-form EP and EPD (see Tab.~\ref{tab:dist_moments}).

{\em Observed shapes and implications.}---Fig.~\ref{fig:dist_hist_twoqubit} shows the histograms for CNOT, $B$, and $\sqrt{\mathrm{SWAP}}$. In these examples, the distributions are unimodal but exhibit different skewness and widths. In particular, CNOT and $B$ gate have the same EP (identical mean entanglement), yet their histograms differ visibly: the $B$ gate is more tightly concentrated, consistent with its smaller EPD, whereas CNOT places more weight on weakly entangling inputs. The $\sqrt{\mathrm{SWAP}}$ gate shows a pronounced peak near small $e$ together with a broad support, illustrating that substantial input-state sensitivity can coexist with a moderate average entangling power. These distribution-level distinctions motivate EPD as a necessary complement to EP: while EP captures the typical entangling strength, EPD captures how uniformly that the strength is realized.

\twocolumngrid

\end{document}